\documentclass[superscriptaddress,preprint,amsmath,amssymb]{revtex4}
\usepackage{graphicx}  
\usepackage{latexsym}
\usepackage{dcolumn}   
\usepackage{bm}        
\usepackage{amssymb}   
\usepackage{textcomp}
\usepackage{amsmath}
\usepackage[dvips]{color}
\usepackage{soul}
\usepackage{float}
\begin{document}
\title{Overstretching of B-DNA with various pulling protocols: Appearance of structural polymorphism and S-DNA}
\author{Ashok Garai}
\email[Corresponding author: ]{ashok.garai@gmail.com}
\address{Centre for Condensed Matter Theory, Department of Physics, Indian Institute of Science, Bangalore 560012, India}
\affiliation{Department of Physics, The LNM Institute of Information Technology, Jamdoli, Jaipur 302031, India},
\author{Santosh Mogurampelly}
\author{Saientan Bag}
\author{Prabal K. Maiti} 
\affiliation{Centre for Condensed Matter Theory, Department of Physics, Indian Institute of Science, Bangalore 560012, India}


\begin{abstract}
We report a structural polymorphism of the S-DNA when a canonical B-DNA is stretched under different pulling protocols and provide a fundamental molecular understanding of the DNA stretching mechanism. Extensive all atom molecular dynamics simulations reveal a clear formation of S-DNA when the B-DNA is stretched along the $3'$ directions of the opposite strands (OS$3$) and is characterized by the changes in the number of H-bonds, entropy and free energy. Stretching along $5'$ directions of the opposite strands (OS$5$) leads to force induced melting form of the DNA. Interestingly, stretching along the opposite ends of the same strand (SS) leads to a coexistence of both the S- and melted M-DNA structures. We also do the structural characterization of the S-DNA by calculating various helical parameters. We find that S-DNA has a twist of $\sim 10$ degrees which corresponds to helical repeat length of $\sim 36$ base pairs in close agreement with the previous experimental results. Moreover, we find that the free energy barrier between the canonical and overstretched states of DNA is higher for the same termini (SE) pulling protocol in comparison to all other protocols considered in this work. Overall, our observations not only reconcile with the available experimental results qualitatively but also enhance the understanding of different overstretched DNA structures. 
\end{abstract}

\maketitle

\section{Introduction}
Double stranded DNA, a semi-flexible molecule \cite{bustamante94,marko95,bustamante97} can be found in different conformational forms \cite{alberts08} both in {\it in vitro} and {\it in vivo} environment. Presence of external force changes its structure and allows to assess the eventual polymorphism exhibited through various conformations with a change in the known helical parameters\cite{dehez17, dans12}. Within the context of the present work we use `polymorphism' to broadly refer to the fact that for a given DNA sequence different pulling protocols give rise to different DNA structures. Single molecule pulling experiments (SMPE) \cite{smith92,bustamante96, caron96,nagy08, williams10} and simulation studies \cite{lavery96,caron96,kosikov99,harris05,piana05,lohikoski05,piana07,roe09,maiti09,harris10,kubar11} revealed that the canonical form of double stranded DNA (B-DNA) stretches almost twice its original contour length when subjected to stretching forces. Experimental studies \cite{smith92,bustamante96,caron96,block97,prentiss03,prentiss08,prentiss09,perkins12,yan12} report an overstretched transition of B-DNA when pulled with force ranging between $60-70$ pN, and the double strands separate at sufficiently high forces \cite{rief99}. A characteristic plateau has been observed in the force-extension curve of dsDNA that signifies a highly cooperative transition from B-DNA to an overstretched conformation \cite{caron96}. The study of the overstretching transition of DNA is essential to understand the DNA elasticity and the interactions of stretched DNA and small molecules.

Two different physical pictures has been emerged to explain the overstretching transition of DNA. Some experiments \cite{bustamante96, caron96, niklas12,karplus14,bolonick96} suggest that the dsDNA adopts a new elongated double stranded structure (called as S-DNA) in the overstretching regime. The stretching curve of such overstretching transition continues to very high forces in comparison to the overstretching force plateau. It has been proposed that in S-DNA base stacking is disrupted but base pairing remains preserved. In the absence of a full microscopic picture of S-DNA structure and its characteristics, the experimental data is often fit to some free parameters to interpret the existence of S-DNA \cite{cocco04,storm03}. In another picture a force induced melting (FIM) of DNA is proposed on the basis of thermodynamic measurements \cite{bloomfield101, peterman09}. Force induced melting occurs mostly in low GC content regions and only in DNA ends or nicks. Whether it is the emergence of S-DNA or the melted DNA in the overstretching regime, in both the cases the nature of resulting DNA structure is understood to depend on the rate of pulling \cite{gaub08}, direction of applied force \cite{lavery96}, salt concentration, DNA sequence and temperature \cite{yan10}.  The structure of the overstretched state of the DNA is not yet fully resolved and is under intense debate \cite{bustamante96, caron96, cocco04,storm03, harris05, bloomfield101, leger99, bloomfield201,peterman09,mishra15} over the last two decades. Thus it is still an open question whether the overstretching transition leads to FIM or gives rise to S-DNA. Various experimental results and theoretical models exist to support both the physical pictures of the overstretched states. In the context of overstretched transition of DNA, Chaurasiya et al. \cite{williams10} in their review article report that FIM of DNA as the only valid picture instead of S-DNA picture because it requires extra fitting parameters and fails to show testable results. This view has been strongly criticized by Krichevsky \cite{krichevsky10} and Whitelam \cite{stephen10}. In their comments they provided supportive arguments in favor of S-DNA and exhibit some weakness of force induced melting picture. As a result McCauley {\it et al.} \cite{micah10} in their `reply to comment' added both the model pictures associated with the overstretching transition of DNA without a concrete conclusion. 

The H-bonded base-pairs in B-DNA prefer tilted orientation with respect to the backbone which prompted Lavery and Lebrun \cite{lavery96} to propose that the nature of overstretched structure depends on pulling scheme. Explicitly, the stretching along $5'$ directions of the opposite strands (OS$5$) is expected to increase the base-pair tilt compared to the stretching along the $3'$ directions of the opposite strands (OS$3$) direction (see Fig.~\ref{fig1-pull-prot} for a schematic). Consistent with the above arguments, Albrecht et al. \cite{gaub08} reported that the dsDNA is more stable when stretched with pulling protocol OS$3$ compared to that employed OS$5$ protocol. However, such differences are seen to be marginal when the rate of pulling force is low. Using constant force SMPE on $\lambda$-phage DNA, Danilowicz et al. \cite{prentiss09} reported that DNA becomes more stable with OS$3$ pulling than with OS$5$ pulling and further pointed out the differences between structures obtained from two different pulling protocols. They observed a narrow backbone spacing and rotated bases for the OS$5$ overstretched structure. Further, they found that the force induced separation of the DNA into single strands occur at higher forces for OS$3$ pulling than for OS$5$ pulling. Depending on the temperature and salt concentration, two types of overstretching transitions have been identified \cite{yan10}. Such transitions in the overstretching regime are characterized by hysteretic and non-hysteretic behavior in the stress-strain curves. In hysteretic double helix reorganization, slow peeling of one strand from its complementary strand occurs, and in non-hysteretic situation, the DNA adopts an elongated double-stranded form \cite{yan10}. Thermodynamic variables associated with these transitions were found to be different \cite{yan12}. For topologically closed but rotationally unconstrained DNA, Paik and Perkins \cite{perkins12} showed that the DNA exhibits overstretching transition at $65$ pN with no hysteresis. They further observed that degree of hysteresis in overstretching transition depends on the number of nicks present in the DNA. However, appearance of ss-DNA in overstretching does not necessarily originate from the free ends or nicks rather peeling is the basic cause for hysteresis in the force-strain curve. In another study, Zhang et al. \cite{yan14} reported interconversion between three overstretched states (S-DNA, DNA bubble, and peeled ssDNA) in the presence of monovalent salt at constant force.

A very few MD studies on the DNA overstretching have been reported in literature \cite{kosikov99, piana05, lohikoski05, roe09, harris10, kubar11, maitis11, maitij15, maiti16}, which predicted a ladder like structure in the overstretching regime. For instance, Balaeff et al. \cite{beratan11} observed a B-to-zip DNA transition instead of B-to-S DNA transition using steered molecular dynamics simulations. They proposed that S-DNA could be an intermediate state during this transition. Coarse-grained DNA models \cite{cieplak09, onufriev13, swigon12, kaxiras12, louis13} have also been used  to understand several mechanisms related to the stretching behavior of DNA. For example, Romano et. al. \cite{louis13} using ox-DNA reported a detailed study of the overstretching transition of dsDNA and showed temperature dependence of overstretching force in good agreement with the available experimental results. However, in their study they did not find any evidence of the S-DNA. Although a qualitative understanding of different overstretching structures is apparent, a clear molecular level evidence and characterization of the S-DNA structure is still not available. Motivated by the above discussed lacunas, we seek to address several important questions regarding the emergence of S-DNA during DNA stretching in this article: (i) Does S-DNA adopt a helical or ladder-like structure? (ii) How does the helical parameters of S-DNA compare to the B-DNA? (iii) Whether the structure attributed to be the S-DNA bear intact base-pairs or internally melted? (iv) How does the potential of mean force (PMF) varies for various overstretching structures? (v) What is the melting force of different overstretched forms? (vi) How do the entropy and enthalpy change during the overstretching transition of DNA? etc. Using fully atomistic MD simulations we provide a comparative study that helps in characterizing the structural changes that occur during the pulling of B-DNA.

\section{Methods}
We employ both non-equilibrium and equilibrium simulations as described in the following: (i) Non-equilibrium simulations: We used AMBER$10$ suite of programs \cite{ucsf} with the AMBER parm99 force field \cite{wang00} along with the parmbsc0 corrections \cite{perez07} (together, referred to as ff10) to describe inter and intra-molecular interactions involving DNA \cite{hornak06}, the TIP3P model \cite{klein83} for water and Joung-Cheatham parameter for ions \cite{joung}. The starting structures for the ds-DNA with the sequences of $12$ base-pairs {\it d(CGCGAATTCGCG)} and $30$ base-pairs {\it d(CGTTGGTGCGGATATCTCGGTAGTGGGATA)} were built using the nucgen module \cite{pearlman95} of the AMBER$10$ suite of programs. Using the LEaP module in AMBER$10$, the ds-DNA structure was solvated with a water box. The box dimensions were chosen sufficiently large enough to ensure a minimum of $10$ {\AA} solvation shell around the DNA in all the three directions in fully stretched form. We used periodic boundary conditions in all three directions during the simulation. To perform non-equilibrium simulations in the framework of constant-force ensemble, we modified the SANDER program \cite{pearlman95} to include the external stretching force \cite{maiti09}. The external force was applied on the O$3'$ and O$5'$ atoms of the two opposite ends of the DNA as schematically described in Fig. \ref{fig1-pull-prot}. Starting from $0$ pN, the external force was increased linearly with simulation time at a stretching rate of $0.0001$ pN fs$^{-1}$.  During the simulation, bond lengths involving bonds to hydrogen atoms were constrained using SHAKE algorithm \cite{ryckaert77}. All the structures of DNA were visually inspected using VMD \cite{schulten96}. (ii) Equilibrium simulations: Umbrella sampling (US) simulations \cite{torrie77} were performed using PMEMD module of AMBER $14$ \cite{case14} with the ff$10$ force field \cite{hornak06}. The PMF was calculated by restraining the force centers with a harmonic potential, $k(r-r_0)^2$, where $k$ is the spring constant, $r$ and $r_0$ are the instantaneous and equilibrium distance between the force centers, respectively. The values of $k$ and separation between successive windows used in the PMF calculations are $4$ kcal/mol-{\AA}$^2$ and $1$ {\AA}, respectively. The reaction coordinate was chosen as the distance, $r$ between force centers at the DNA termini. The choice of the force constant and the sampling window size was motivated by our previous US simulation studies \cite{maitim13}, which showed good convergence. The sampling of 1-D reaction coordinate was varied with a separation of $1$ \AA{} to capture the entire unstretched canonical form to fully stretched conformation of DNA, ranging from $38${\AA} to $94${\AA} and $38${\AA} to $100${\AA} for OS$5$ and OS$3$ stretching, respectively. In each window $1$ ns NPT (constant number, constant pressure, and constant temperature, respectively) and $1$ ns NVT (constant number, constant volume, and constant temperature, respectively) production runs were performed \cite{frenkel02}. We used the equilibrated structure of the previous distance window as the starting structure for the subsequent window. Finally, weighted histogram analysis method \cite{kollman92, grossfield} was used to calculate the unbiased PMF by subtracting the contribution of the biased harmonic potential. We have further verified (see SI: section I) that $1$ ns long trajectory gives sufficiently converged PMF (see figure S9 in SI). To check the convergence of PMF, we also continued our production runs up to $30$ ns in each distance window for OS$3$ and OS$5$ cases (see Fig. S10). All the helical parameters such as the rise, roll, shift, slide, tilt, and twist of the DNA were calculated using the Curves+ suite of programs \cite{lavery09}. The remainder of the paper is organized as follows:  In the section III we discuss the various results obtained from our non-equilibrium and equilibrium pulling simulation. Section III A discusses the behavior of force-strain curves, section III B describes the effect of force on hydrogen bonds and emergence of S-DNA, section III C introduces the segmental analysis of H-bonds, section III D provides thermodynamics of different over- stretched states associated with various pulling scheme and finally section III E includes the effects of DNA length of S-DNA emergence. We note that the sections III A-C and III E discuss the results obtained from the non-equilibrium simulations and section III D introduces the results obtained from the equilibrium simulations.

\section{Results and Discussion}

\subsection{Force-strain curves} 
\begin{figure}
\centerline{\includegraphics[width=1\linewidth]{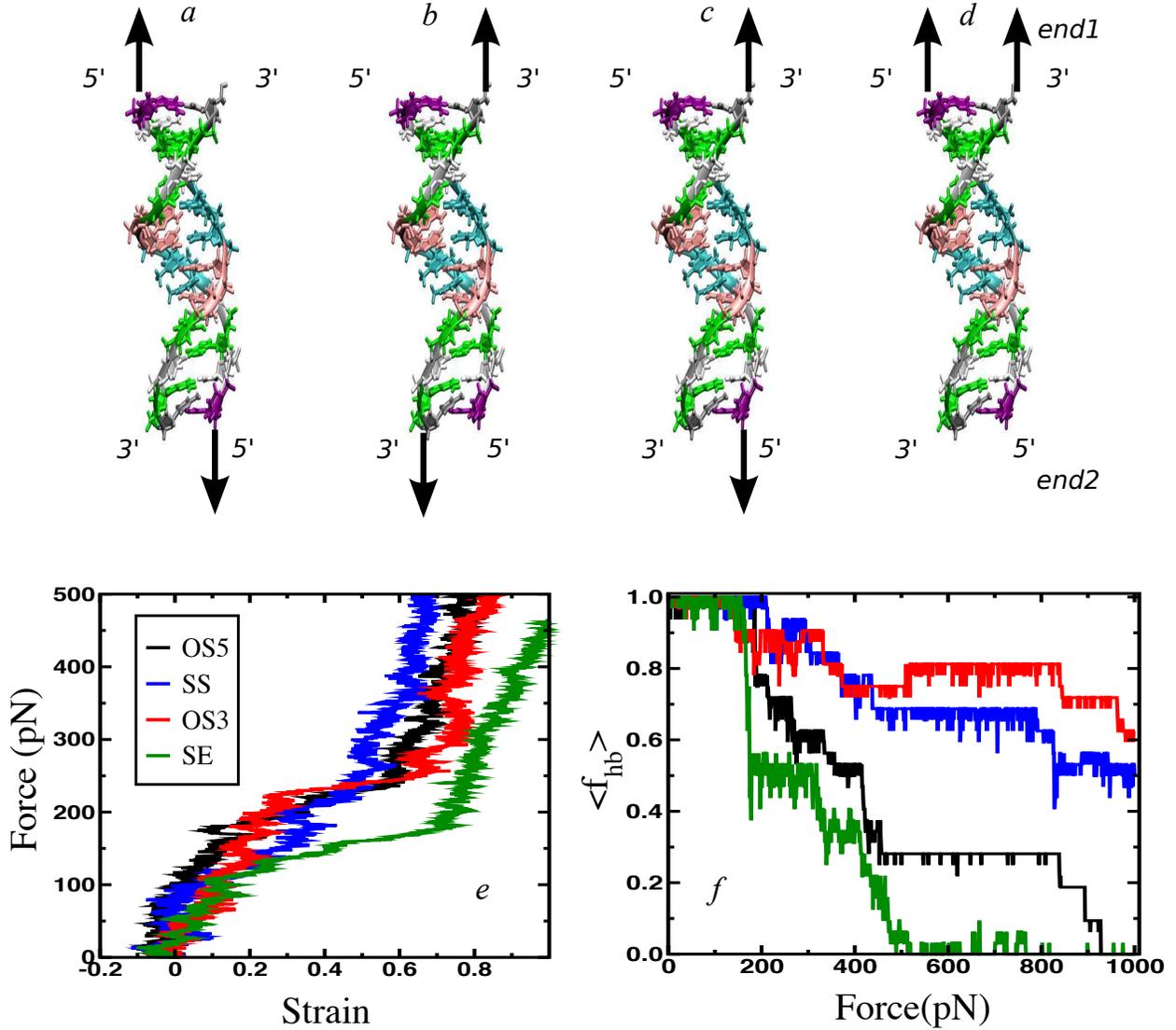}}
\caption{First row: Schematic describing different pulling protocols considered in this work: (a) pulling $5'$ends of the opposite strands (OS$5$), (b) pulling $3'$ends of the opposite strands (OS$3$), (c) pulling opposite ends of the same strand (SS), and (d) pulling both strands at the same end (SE). Arrows indicate the direction of the applied force. For case (d) we kept end $2$ fixed while pulling the other end of the DNA. For describing various pulling schemes, we defined $5'3'$ end and $3'5'$ end of the ds-DNA as end$1$ and end$2$, respectively. Second row: (e) Force strain behavior and (f) $<f_{hb}>$ as a function of applied force for various pulling protocols. For the OS$3$ pulling protocol, almost $70\%$ H-bonds remain intact even when the DNA is fully stretched leading to the emergence of S-DNA.}
\label{fig1-pull-prot}
\end{figure}

We apply external force on $O3'$ and $O5'$ terminal atoms of the two strands of the B-DNA in four different ways as schematically illustrated in Fig. ~\ref{fig1-pull-prot}. The strain is defined as the ratio of the instantaneous change in DNA length (taken as the average end-to-end $O3'$-$O5'$ length of two strands) to its initial contour length \cite{garai15}. The force versus strain curves obtained from non-equilibrium pulling simulations are displayed in Fig. \ref{fig1-pull-prot}(e). Force-strain behavior is widely used to understand the DNA structure and dynamics. For all the present cases, we observe plateau and overstretching regimes, consistent with typical force-strain features reported previously \cite{bustamante96,prentiss09,maiti09}. However, we note that the length of the $\lambda$-DNA considered in experiments span few kilo base-pairs and the rate of pulling is slow compared to our atomistic simulations. This leads to the differences observed in force-extension behavior at a quantitative level. We find that at extremely low forces, the structural changes of the B-DNA are entropy driven, which can well be described by the inextensible worm-like chain model \cite{bustamante97,marko95,bustamante94}. In this regime, the extension of ds-DNA beyond its initial contour length is negligible. Further pulling leads to a plateau region where the DNA is observed to stretch about 1.6-1.7 times its initial contour length with a small increase in applied force. Interestingly, such plateau region was not observed in earlier experimental studies for short ($< 30$ bp) DNA pulling \cite{strunz99, pope01}. We further observe that the plateau width varies depending on the pulling scheme. At large forces, the base-pair to base-pair stacking potential can no longer stabilize the B-form configuration of ds-DNA resulting in a structural transition from the canonical B-form to a new overstretched conformation. At such large forces, the traditional WLC model is inadequate to understand the force-extension behavior \cite{bustamante97,marko95,bustamante94}. 

\subsection{Hydrogen bond breaking and emergence of S-DNA} We further examined the number of intact H-bonds (HBs) which is considered to be a fundamental parameter in characterizing various forms of DNA conformations. Explicitly, it is known that the interaction potential emanating from the formation of HBs between base-pairs stabilizes the canonical B-DNA structure, and consequently the breaking of HBs gives rise to a melted DNA \cite{ferrira15}. Thus, the number of HBs is an important quantity in monitoring conformational changes that occur during the pulling of B-DNA. We define fraction of HB $(f_{hb})$ at a given force as the ratio of intact HBs to the maximum possible HBs in the B-DNA. In our calculation, a hydrogen bond is defined to be formed between the donor, D and the acceptor, A provided that the distance between D-A is less than the cutoff distance $2.7${\AA} and the angle D-H-A is greater than $130^\circ$. Fig.~\ref{fig1-pull-prot}(f) displays $f_{hb}$ as a function of the external force for different pulling protocols. Strikingly, we find that $f_{hb}$ depends strongly on pulling protocol employed. Interestingly, we observe that while almost $75\%$ of the H-bonds are {\em broken} with OS$5$ pulling protocol in the fully stretched conformations, almost $70\%$ H-bonds remain {\em intact} with OS$3$ pulling scheme. Similarly, while almost all H-bonds are broken with SE pulling, only less than half H-bonds are broken with SS scheme.

\begin{figure}
\centerline{\includegraphics[width=1\linewidth]{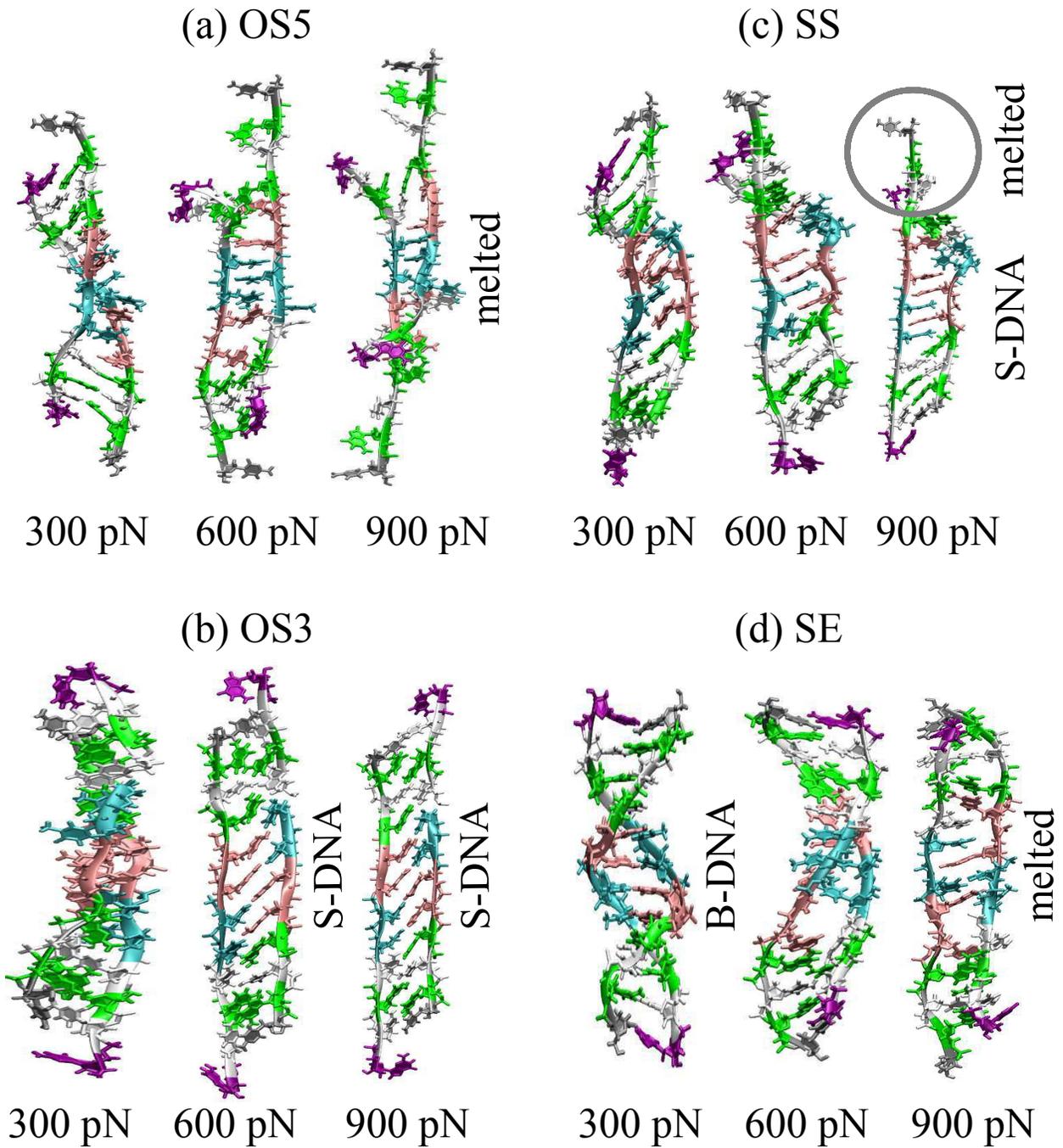}}
\caption{Instantaneous snapshots of the DNA at different pulling forces for different pulling protocols: (a) OS$5$, (b) OS$3$, (c) SS and (d) SE.}\label{fig2-snap-ne}
\end{figure}

Structural snapshots of the stretched DNA obtained from different pulling protocols are displayed in Fig.~\ref{fig2-snap-ne} to visually aid in understanding the degree of H-bond breaking at a qualitative level. From the snapshots and broken H-bonds, it is apparent that dsDNA completely melts under the application of external force via OS$5$ scheme (see Fig.~\ref{fig2-snap-ne}(a)). More interestingly, S-DNA bearing most of the intact H-bonds is seen to emerge in the overstretched regime when the canonical DNA is subjected to OS$3$ pulling protocol (see Fig.~\ref{fig2-snap-ne}(b)). However, with further increase in applied force, the S-DNA eventually transforms into melted DNA as expected. This also indicate that the force required to obtain melted DNA is large when OS$3$ protocol is employed compared to the OS$5$ pulling protocol. Interestingly, the snapshots shown in Fig.~\ref{fig2-snap-ne}(c) and H-bonds presented in Fig.~\ref{fig1-pull-prot}(f) suggest a coexistence of S-DNA and melted DNA with SS pulling protocol. On the other hand, similar to the results obtained with OS$5$ scheme, a complete melted DNA is obtained with SE pulling protocol (see Fig.~\ref{fig2-snap-ne}(d)).

Overall, the above results suggest that when the DNA is pulled through OS$3$, only about $20\%$ of HBs are cleaved and consequently a clear B-to-S structural transition emerges.

\subsection{Nature of H-bond breaking} 
\begin{figure}
\centerline{\includegraphics[width=1\linewidth]{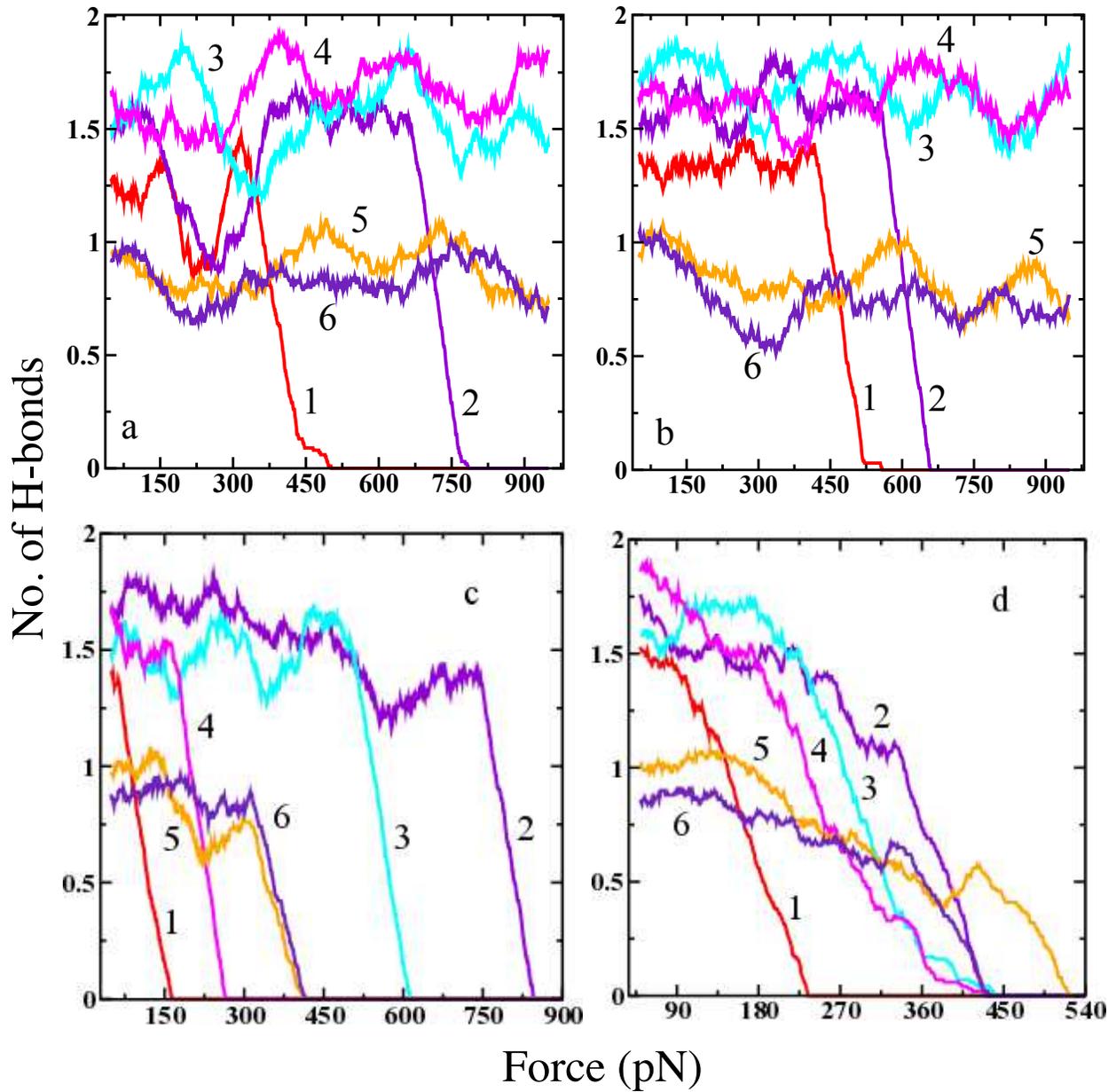}}
\caption{Number of H-bonds as a function of applied force for OS$3$ (first row) and OS$5$ (second row) pulling cases for two independent simulations. The legend shown in the figures indicate the position of the base-pair between which H-bonds are being observed; $Ô1Õ$ denote terminal base-pairs, $Ô2Õ$ denote the base-pair next to it and so on. In both the initial conditions, the terminal H-bond breaks first. Rest of the H-bonds remain intact up to $1000$pN for OS$3$ case whereas for OS$5$ case breaking events of H-bonds are stochastic in nature.}\label{fig3-hb-33-55}
\end{figure}

Now we ask the following questions: which HBs in the DNA are most vulnerable under the application of force? Are the H-bond breaking events deterministic or stochastic? Towards the above objectives, we repeat the non-equilibrium stretching simulations with different initial conditions and calculate the number of H-bonds between the individual base-pairs starting from the terminal end as a function of applied force. The results for two independent OS$3$ pulling simulations are shown in Fig. \ref{fig3-hb-33-55}(a-b). In both the cases, H-bonds at the terminal base-pair (G-C) is seen to break before any other H-bonds are affected. The results presented in Figs. \ref{fig3-hb-33-55}(a-b) indicate that the breaking of H-bonds occur in the vicinity of terminus (labeled as position $1, 2$ in Figs. \ref{fig3-hb-33-55}(a-b)), where stronger G-C base-pairing is located, despite the presence of weaker A-T segments in the neighbor regime. More interestingly, the H-bonds far from the force center are seen to be intact including the AT rich segments, even at higher destabilizing forces. The above result is arguably a consequence of the cooperative nature of DNA stretching. Further, these results suggest a weakening of terminus regime under the influence of applied force, which is also consistent with the reports of Ferreira et al. \cite{ferrira15}. Additionally, segmental analysis of H-bonds with different pulling rates are presented in Figure S1. Similar results are observed with different pulling rates and initial configuration revealing the deterministic nature of H-bond breaking at the terminus and rationalizing the above arguments. Overall, the results presented in Figs. \ref{fig1-pull-prot} and  \ref{fig3-hb-33-55}(a-b) along with the snapshots shown in Fig.~\ref{fig2-snap-ne}(b) conclusively demonstrates the emergence of S-DNA when a canonical B-DNA is subjected to stretching via OS$3$ protocol.

We similarly repeat the non-equilibrium stretching simulation for OS$5$ with a different initial condition and calculate the number of HBs between the individual base-pairs starting from the terminal end as a function of applied force (see Fig.~\ref{fig3-hb-33-55}(c-d)). In accordance with the OS$3$ pulling case, the H-bonds at the terminal base-pair (G-C) are seen to break first while the subsequent cleaving events are completely stochastic. Specifically, in one initial condition, the terminal H-bonds break first followed by $4$th (G-C), $5$th (A-T) and $6$th (A-T) respectively while in the other case the respective breaking events sequence is $1$st (G-C), $4$th (G-C), $6$th (A-T) and $2$nd (G-C). Initially, when all H-bonds in DNA are intact in the stable form, the propagation of applied force along the DNA backbone follow a specific manner. As the H-bond are cleaved, the corresponding free bases attain more degrees of freedom and subsequently the DNA becomes more destabilized. Eventually, the force profile along the backbone is more randomly distributed leading to uncorrelated cleaving events.   

\subsection{Insights from equilibrium simulations} So far we have discussed the appearance of DNA structural polymorphs and emergence of S-DNA using non-equilibrium stretching simulations. To gain further insights into the stretching phenomena, we perform additional simulations by introducing an Harmonic potential to held the DNA end-to-end length at a desired value so as to mimic DNA stretching in an equilibrium manner. The equilibrium simulations are also expected to eliminate the artifacts introduced by the faster pulling rate, low sampling of overstretched states and therefore can be used to rationalize the observations made using the non-equilibrium simulations. 

\begin{figure}
\centerline{\includegraphics[width=1\linewidth]{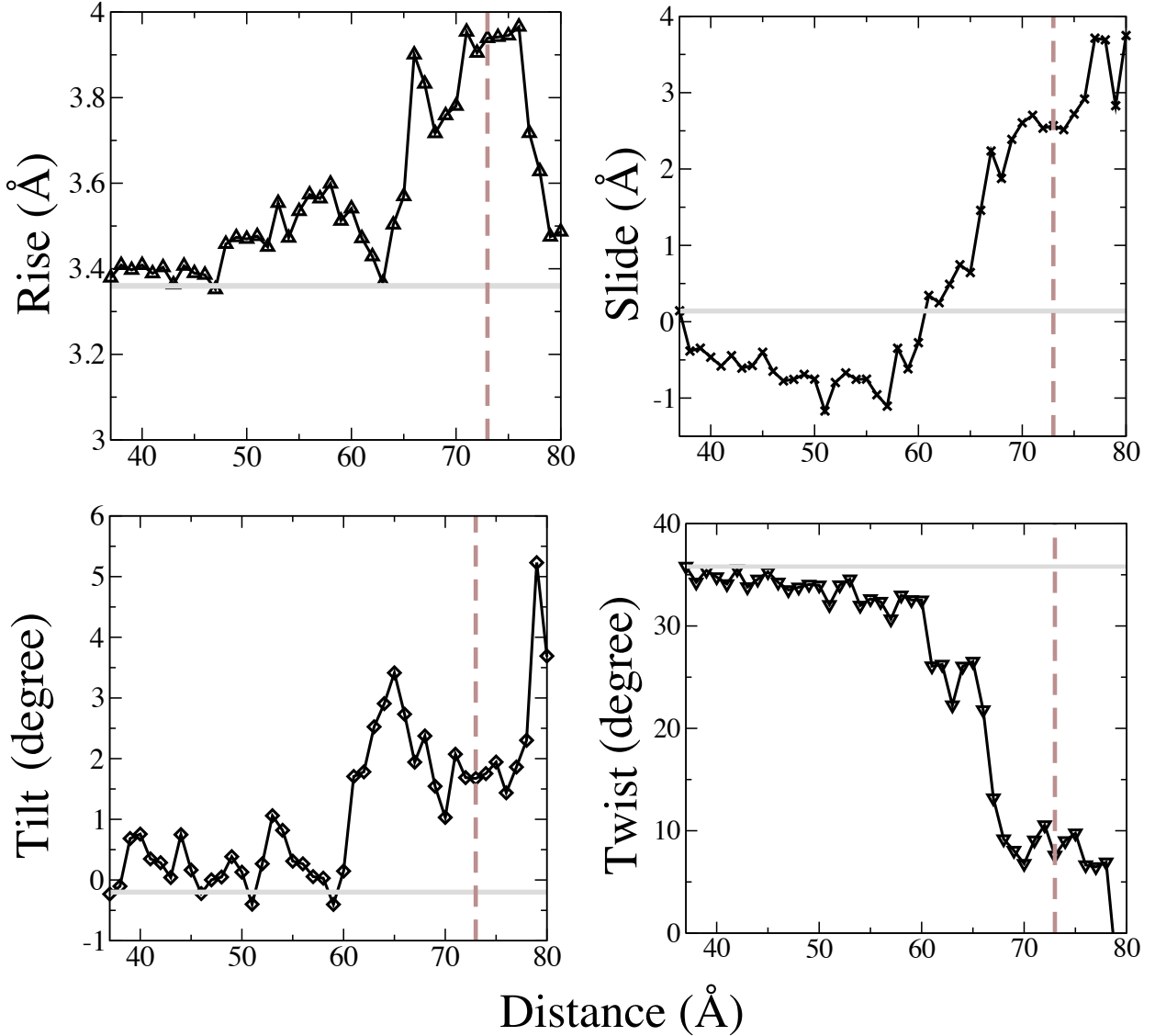}}
\caption{Average inter base-pair parameters of S-DNA are obtained from OS$3$ equilibrium pulling data. The average inter base-pair parameter for B-DNA are as follows: Rise$=3.36$; Slide$=0.14$; Tilt$=-0.2$; Twist$=35.8$; \cite{lavery09} and are represented by a solid horizontal line in each plot for reference. Right side of the vertical dotted line (corresponds to $ \sim 140$ pN, obtained from equilibrium stretching curve) in each plot refers to the helical parameters of the emerging S-DNA.}\label{fig4-hel-parm}
\end{figure}

To characterize the structure of S-DNA to full extent, we considered the Harmonic restraining potential between the OS$3$ oxygen atoms, for which the emergence of S-DNA is predicted by non-equilibrium simulations. Various inter base-pair helical parameters of DNA at different equilibrium end-to-end lengths of the Harmonic potential are displayed in Fig.~\ref{fig4-hel-parm} for OS$3$ pulling case. We find that the rise, slide and tilt increases with equilibrium OS$3$ length with respect to its canonical B-DNA (see the reference line) \cite{lavery09}. Interestingly, the twist is seen to decrease with respect to its canonical B-DNA and saturates to a value of $\sim10$ degrees. This corresponds to a helical repeat of $36$ bp, which agrees very well with the previous experimental results of $\sim 38$ bp \cite{leger99}. The other helical parameters are shown in Fig.~S2 and in Fig.~S3. Our study further shows that stretch, stagger, and propel increase and shear, and opening decrease in comparison to the canonical B-DNA values. Similarly, comparison to inter base-pair parameters of S-DNA indicates that rise, and roll increase and shift, and twist decrease. 

\begin{figure}
\centerline{\includegraphics[width=1\linewidth]{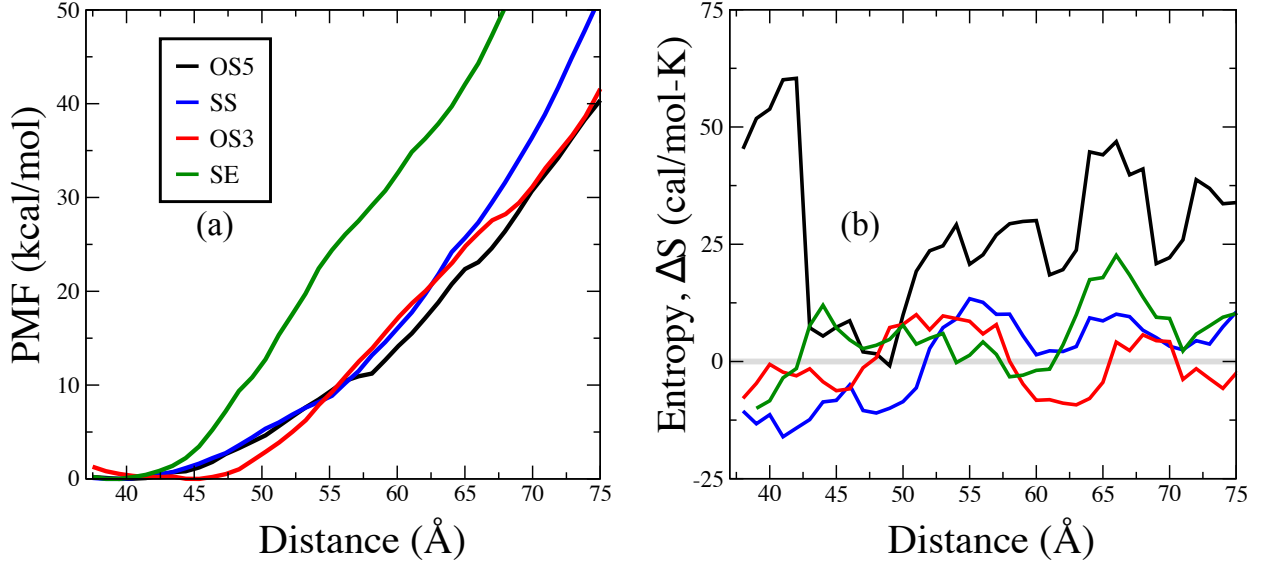}}
\caption{(a) The PMF and (b) Change in entropy as a function of stretching length as obtained from the equilibrium simulations with running average of 5 data points.}\label{fig5-pmf-entropy}
\end{figure}

We further calculated the potential of the mean force (PMF) by accumulating the histograms of end-to-end distances for different end-to-end oxygen atom references. The PMF can further provide information on the dynamics of DNA folding/unfolding phenomena. In Fig.~\ref{fig5-pmf-entropy}(a), we present results of PMF for various pulling scenarios obtained using the Umbrella sampling (US) framework. We find that the PMF is higher for SE pulling in comparison to all other cases of pulling. Interestingly, for OS$3$ pulling case, the PMF is little higher than the OS$5$ pulling case at the stretching distances where we have observed S-DNA. We also find appearance of step like features at larger stretching distances ($60$ {\AA} and above). A closer look at the DNA structure reveals that above $60$ {\AA} stretch length,  DNA retains its B-DNA feature before it adopts the S-DNA structure. A structural change causes such drop in force which also leads to a step like shape in PMF for OS$3$ above $60$ {\AA} (see Fig. S11). To understand the contribution of entropy (and equivalently the enthalpy) to free energy, we employed two-phase thermodynamic (2PT) method developed by Lin et al. \cite{goddard03,goddard10,goddard11} and calculate the entropy of DNA during overstretching processes. Fig.~\ref{fig5-pmf-entropy}(b) depicts the change in total entropy of DNA at various equilibrium distances for different pulling protocols. For OS$5$ case we find DNA becomes relaxed around $40$\AA{} which invokes a sharp increase in entropy but for other cases the DNA stiffens initially and then becomes relatively more flexible. This is reflected in the trend of change of $\Delta S$ with respect to distance (initially it decreases and then increases). We further notice that for B-to-S transition (OS$3$ pulling case), a negative change in entropy occurs consistent with the recent experimental study by Zhang et al \cite{zhang13}. Interestingly, for OS$5$ case, we observe a positive change in entropy. 

\begin{figure}
\centerline{\includegraphics[width=1\linewidth]{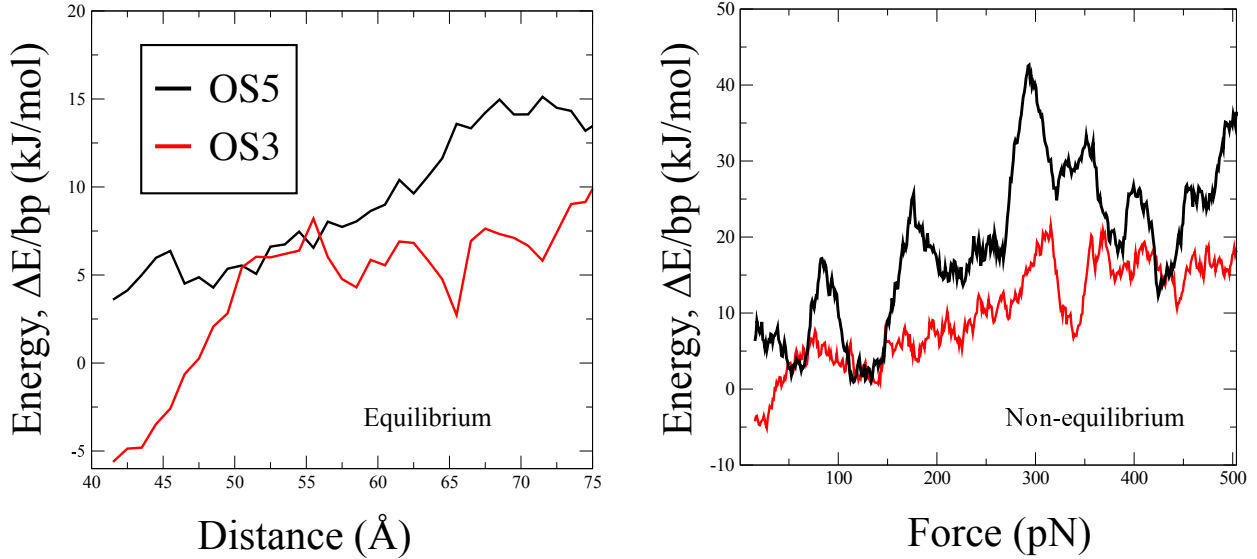}}
\caption{Change in total potential energy, $\Delta E$ (enthalpy) plotted as a function of distance and force for equilibrium and non-equilibrium cases, respectively. The $\Delta E$ values are averaged over $5$ and $30$ data points for equilibrium and non-equilibrium cases, respectively.}
\label{fig6-energy}
\end{figure}

\begin{figure}
\begin{center}
\includegraphics[width=0.95\columnwidth]{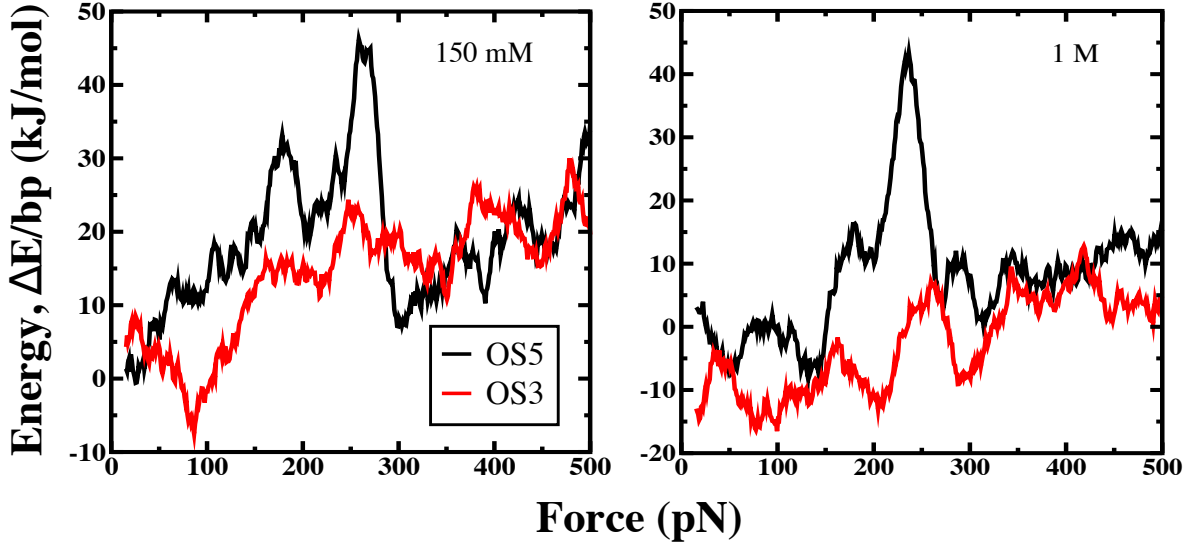}\\
\end{center}

\caption{Change in potential energy ($\Delta E$) as a function of applied force for OS$5$ and OS$3$ nonequilibrium pulling data for $150$ mM and $1$ M salt concentrations.}
\label{fig7-33-55-nep}
\end{figure}

We also extract force-distance curve from our equilibrium simulations for different pulling protocols and display in Fig. S4 (a). We observe irregular variations in force-distance curve displaying local maxima at different equilibrium distances. The peaks of local maxima indicate a force required for breaking different base-pairs of the dsDNA in different pulling protocols. We also observe that force required to break the dsDNA in OS$3$ pulling is more compared to OS$5$ pulling, consistent with the results obtained from non-equilibrium simulations. However, large force is required to melt DNA through SE pulling process. Our US simulations indicate that the DNA melts (defined as a conformation with $80\%$ of the H-bonds are broken, see Fig. S4 (b)) at different stretching forces when pulled with different pulling protocols. More quantitatively, we observe that the melting forces are $198$ pN, $282$ pN, $781$ pN, and $778$ pN when pulled with OS$5$, OS$3$, SS and SE protocols, respectively. The above results are very similar to those reported by Danilowicz et al. \cite{prentiss09} from the SMPE studies. Our results also support the hypothesis made by Lebrun and Lavery \cite{lavery96}. Additionally, the structural snapshots for OS$5$ and OS$3$ cases taken from our equilibrium simulations (see Fig. S5, Fig. S12, Fig. S6 and Fig. S13) also qualitatively agree with the structures obtained from non-equilibrium pulling simulations for the respective cases. We further calculate the changes in the potential energy of the DNA during pulling processes with respect to the un-stretched DNA. Large contributions come from van der Waals, angle and bond interactions (see Fig. S7). Using our equilibrium simulation data we show the change of the potential energy as a function of distance for two different pulling protocols and is displayed in Fig.~\ref{fig6-energy}. Total potential energy change depends on the direction of pulling. We observe in the beginning of pulling a negative change in the total potential energy along OS$3$ pulling direction and a positive change along OS$5$ pulling direction. This observation originates from the tilt orientation of the base-pairs of the B-DNA. In the beginning of pulling for OS$3$ case tilt orientation of base-pairs decreases, which causes a decrease in energy and yields a negative enthalpy change ($\Delta E$). Whereas for OS$5$ pulling case tilt orientation increases which causes a positive enthalpy change. Our equilibrium study is consistent with the non-equilibrium study (see Fig.~\ref{fig6-energy}). We also observe that $\Delta E$ increases further and further as the DNA is stretched. It clearly indicates the destabilization of the helical structure of the DNA. We find that OS$3$ pulling leads to lower potential energy than OS$5$ direction. In the intermediate distances we further observe a plateau in $\Delta E$ (see equilibrium case of Fig.~\ref{fig6-energy}) for OS$3$ pulling case. This indicate a new conformation of DNA (S-DNA) that remains stable for some time during the OS$3$ pulling process. We next observe a decrease in the change of the electrostatic energy and an increase in van der Waals energy with distances (see Fig. S6). In the overstretched structures the negatively charged phosphates stay far apart from each other, which minimizes their mutual repulsion. In the intermediate distances for OS$3$ pulling case, a stacking structure of dsDNA is observed, which indicates a stable short range interactions. In this situation the backbone strands come closer and opposite charges and group dipoles find a favorable arrangements. We further observe that energy changes due to angle (say $\Delta E_{\theta}$) first decreases and then increases for OS$3$ and OS$5$ pulling case. A significant increase in $\Delta E_{\theta}$ is found for OS$3$ pulling case. Energy change due to bond for OS$3$ pulling case is larger than that for the OS$5$ pulling case. We observe an opposite behavior in the change of the dihedral energy (say $\Delta E_d$) for OS$5$ and OS$3$ pulling cases. Interestingly, $\Delta E_d$ for OS$3$ increases initially and then decreases with the distances. Overall, the increasing nature of total potential energy change indicates the destabilization of helical structure of DNA with stretching. Our results for $\Delta E$ agrees well with the earlier studies by Zhang et al. \cite{zhang13}. Additionally, we perform the non-equilibrium pulling simulations for OS$3$ and OS$5$ pulling in $150$ mM and $1$ M salt concentrations and calculate the $\Delta E$ as a function of applied force (see Fig.~\ref{fig7-33-55-nep}). We observe that the qualitative behavior of the change in $\Delta E$ with the applied force remains same as it is found for no salt concentration. 


\subsection{Effects of DNA length on the emergence of S-DNA} 
\begin{figure}
\begin{center}
\includegraphics[width=0.93\columnwidth]{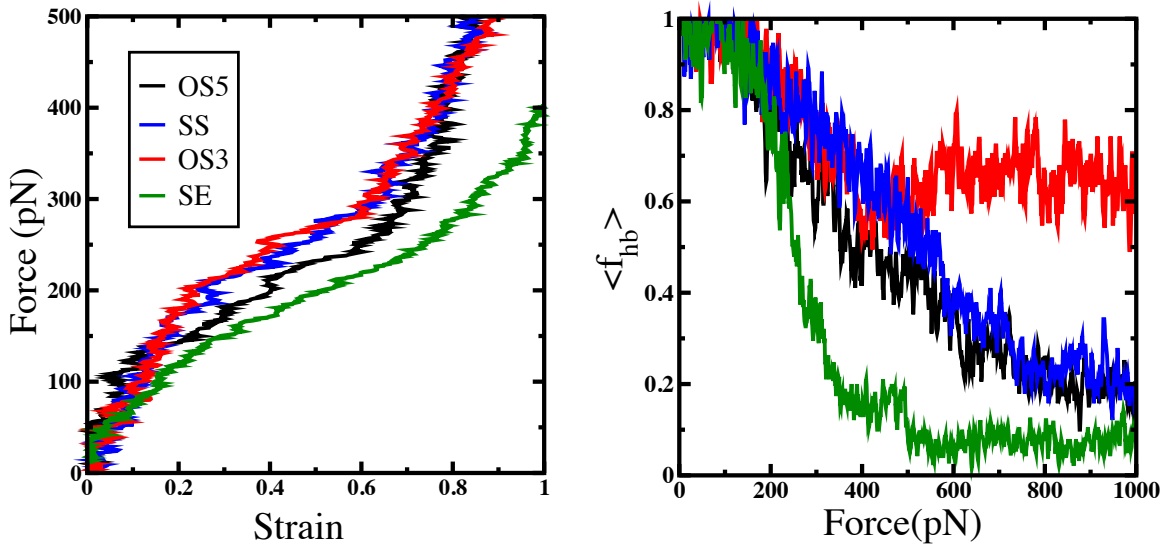}\\
\end{center}
\caption{Force extension behavior and $\langle f_{hb} \rangle$ (fraction of H-bond) of a $30$ base-pair DNA under various pulling protocols. The qualitative features observed for a $30$ base-pair DNA are seen to be consistent with the respective results obtained for $12$ base-pair DNA (see Fig.~\ref{fig1-pull-prot}(e-f))}
\label{fig8-33-55-nep}
\end{figure}

\begin{figure}
\begin{center}
\includegraphics[width=0.95\columnwidth]{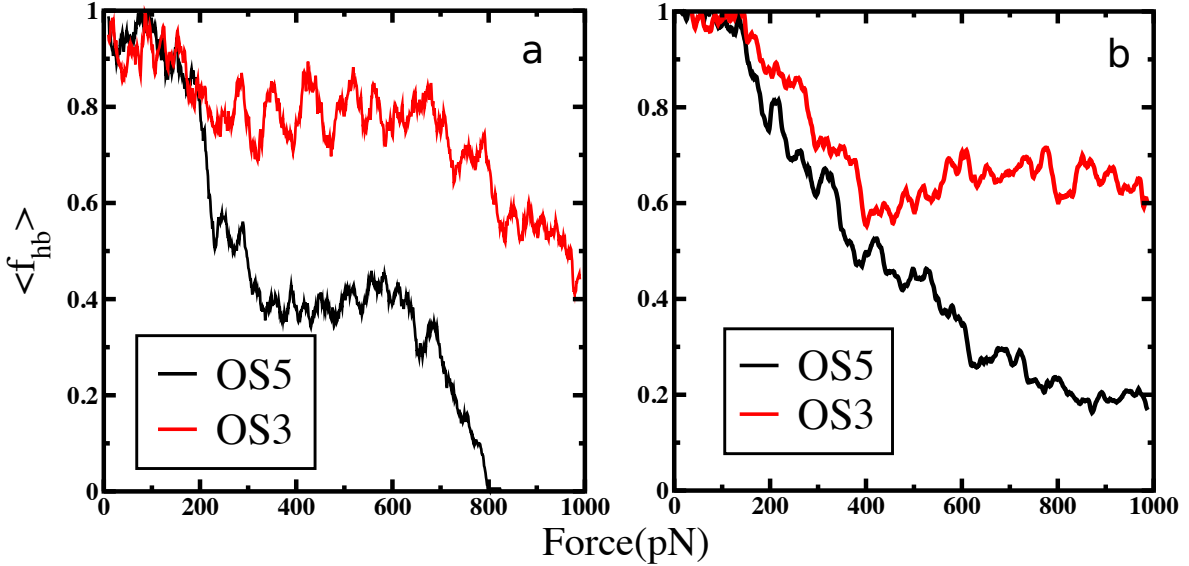}\\
\end{center}
\caption{Effects of DNA length on the fraction of intact H-bonds at $0.0001$ pN/fs pulling rate: (a) $12$ base-pair length and (b) $30$ base-pair length.}
\label{fig9-33-55-nep}
\end{figure}

We note that the results presented so far correspond to a DNA of $12$ base-pairs length under single pulling rate. To check the veracity of the results discussed in this article, we considered stretching of a longer DNA of $30$ base-pairs. The corresponding results are displayed in Figs.~\ref{fig8-33-55-nep} - \ref{fig9-33-55-nep}, which conclusively support the original key observations. In a nutshell, the main conclusions drawn from a single pulling rate are seen to remain intact irrespective of the pulling rates and DNA length. Explicitly, Fig.~\ref{fig8-33-55-nep} depicts the effect of pulling protocol on force-extension behavior as well as H-bond breaking is qualitatively insensitive to DNA length, and Fig.~\ref{fig9-33-55-nep} shows the qualitative features of H-bond breaking under different pulling protocols and consequent emergence of S-DNA (see Fig. S14) with OS$3$ scheme (and melted-DNA with OS$5$ scheme) is independent of DNA length. It is worth mentioning that apart from the $12$ bp and $30$ bp length studied here, we have also looked at the force extension behavior for $20$ bp long dsDNA in our recent work \cite{maitib16}. All these simulations prompted us to comment that the force-extension and H-bond breaking behavior is ÒqualitativelyÓ insensitive to DNA length. However, different DNA sequence (more GC contents for example) may affect the H-bond breaking ÔquantitativelyÕ.

Overall, our equilibrium and non-equilibrium simulations clearly indicate the existence of a complete S-DNA structure when pulled with OS$3$ protocol and a coexistence of both the S-DNA and B-DNA structures when pulled with SS protocol. Using different pulling protocols we also plot (see Fig. S8) force-extension as well as the $<f_{hb}>$ as a function of force for three independent runs at moderate pulling rate, which demonstrates qualitatively similar behavior. The molecular level insights using extensive atomistic simulations reported here shed light on the mechanisms underlying the DNA overstretching transition. 

\section{Conclusions}
In this article, we investigated the effect of stretching force on B-DNA when pulled with different protocols using fully atomistic MD simulations. We provided structural snapshots at the atomistic level adopted by B-DNA under the influence of external force with various schemes providing further insight into the structural aspects. A plateau region of variable widths can also be seen in typical force extension plots for such short DNA pulling. Our simulations suggest that it is possible to obtain a clear B-to-S structural transition when the B-DNA is pulled along OS$3$ and SS direction. We observed that the OS$3$ pulling leads to the emergence of S-DNA whereas the OS$5$ pulling does not give rise to such S-DNA conformation. The possible origin of obtaining S-DNA in OS3 pulling is due to the tilt orientation of the bases with respect to the DNA backbone. OS5 pulling favors the base pair tilt compared to the OS3 pulling. This is consistent with the proposal made by Lebrun and Lavery \cite{lavery96}. From our study, we also realized that the emergence of S-DNA is a more sensitive function of the pulling protocol and the orientation of the base-pairs with respect to the backbone rather than the nature of complementarity of base sequences. Our study further indicates that more than $80$\% hydrogen bonds of the DNA break at higher forces for OS3 pulling in comparison to the OS5 pulling. This observation is similar to that observed earlier in two independent experimental studies \cite{prentiss09,gaub08}. Our results also showed that SS pulling results in a coexistence of S-DNA and melted DNA structure whereas same end pulling does not show such conformation. This behavior is different from the earlier observation by Balaeff et al. \cite{beratan11}. This may be due to the choice of different DNA sequence and pulling protocols. We also compared some of our present results (data not shown) with parm99 force field and found consistent behavior. It will be interesting to see how the present overstretching behavior compares when simulated using a more refined force field parmbsc1 \cite{ivani16}. Further insights into the stretching phenomena and the mechanisms of S-DNA emergence are provided by calculating the PMF, entropy and enthalpy using different simulations setup. We observed a negative change in entropy for B-to-S transition, consistent with the experimental observation \cite{zhang13}. The structural characteristics of S-DNA are investigated in terms of various inter and intra base-pair helical parameters and compared with the standard B-DNA structure. The simulation results are in good agreement with available experimental results \cite{prentiss09, yan12, gaub08, yan10, yan14}. These results may provide more information on active biological processes involving DNA extension and help in understanding the mechanical properties of DNA in a previously unexplored domain. In principle, our new findings and predictions can be tested by carrying out {\it in vitro} single molecule experiments. Overall, our study provides a detailed understanding of the mechanisms underlying S-DNA emergence and structural transitions of dsDNA with different force attachments.

\section{Supplementary Material}
See supplementary material for figures from S1 to S14. These depict various average inter base pair and intra base pair parameters of S-DNA with a comparison to B-DNA at different equilibrium end-to-end lengths; variation of force-distance and fraction of H-bonds with distance from equilibrium simulation data; few snapshots of the DNA from equilibrium simulations for OS$3$, OS$5$ pulling; change of potential energy components for OS$3$, OS$5$ equilibrium simulation; force extension behavior and fraction of hydrogen bond as a function of applied force for various pulling protocols for three independent runs at a moderate pulling rate; discussion on convergence of PMF calculations;  and finally, few snapshots from non-equilibrium simulation data for $30$ bp DNA.

\begin{acknowledgments}
We greatly acknowledge the financial support received from the DAE, India.
\end{acknowledgments}

\end{document}


\title{Supporting Material for ``Overstretching of B-DNA with various pulling protocols: Appearance of structural polymorphism and S-DNA"} 
\author{Ashok Garai}
\email[Corresponding author: ]{ashok.garai@gmail.com}
\address{Centre for Condensed Matter Theory, Department of Physics, Indian Institute of Science, Bangalore 560012, India}
\affiliation{Department of Physics, The LNM Institute of Information Technology, Jamdoli, Jaipur 302031, India},
\author{Santosh Mogurampelly}
\author{Saientan Bag}
\author{Prabal K. Maiti} 
\affiliation{Centre for Condensed Matter Theory, Department of Physics, Indian Institute of Science, Bangalore 560012, India}


\maketitle 

\begin{figure}
\begin{center}
\includegraphics[width=0.98\columnwidth]{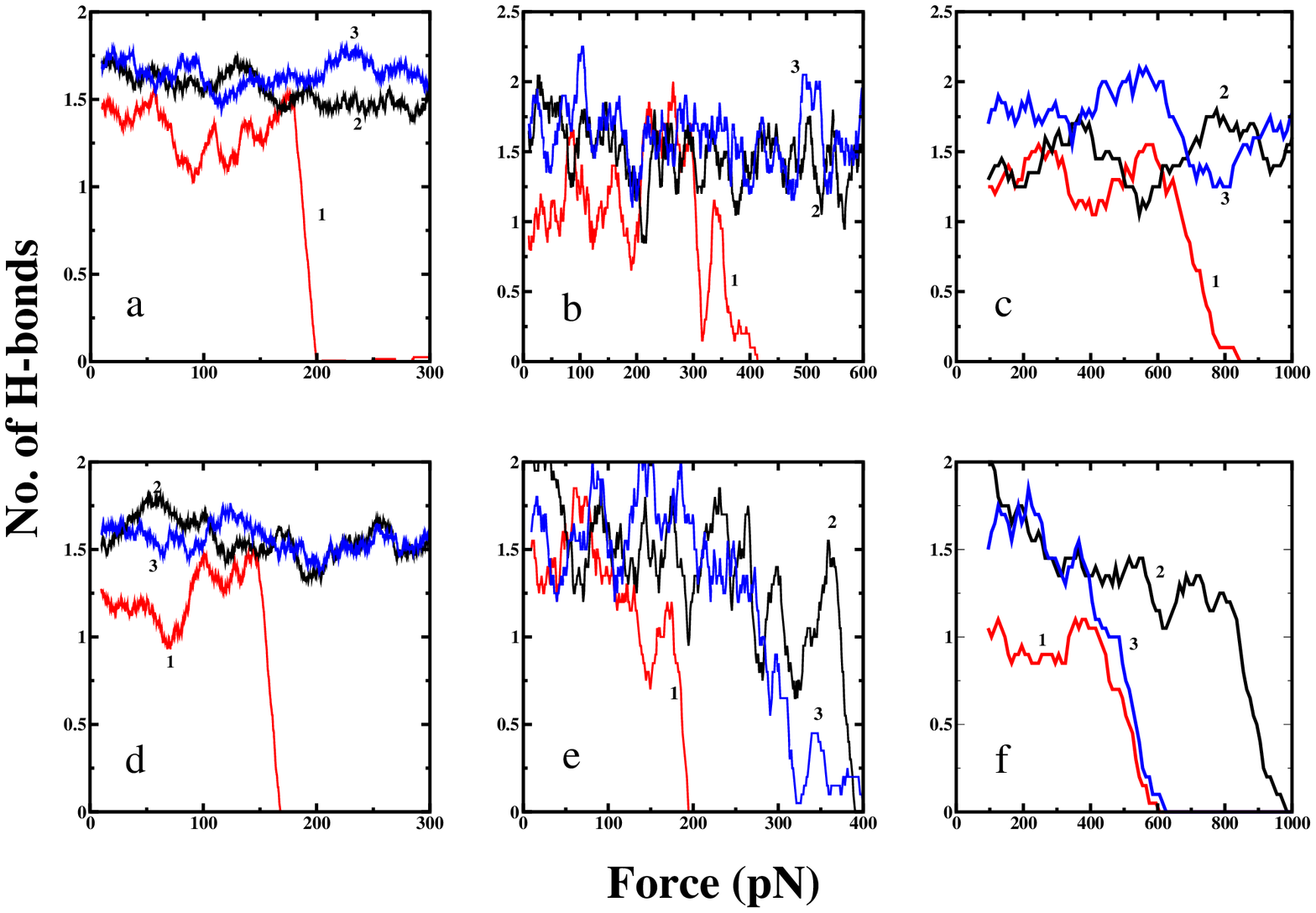}\\
\end{center}
\caption{Effects of pulling rate (each row from left to right: $10^{-5}$ pN/fs, $10^{-4}$ pN/fs and $10^{-3}$ pN/fs) on the segmental H-bond breaking for pulling protocols (a-c) OS$3$ and (d-f) OS$5$. The labelled index denotes the position of the base-pairs from terminus where the breaking of the H-bonds is monitored: Ò1Ó indicate a terminal base-pair, Ò2Ó indicates the base-pair next to it and so on. In all the cases of pulling rates, the emergence of various DNA structural polymorphs is seen to begin with the breaking of H-bonds at the terminus.}
\label{figs1-33-55-nep}
\end{figure}
\begin{figure}
\begin{center}
\includegraphics[width=0.95\columnwidth]{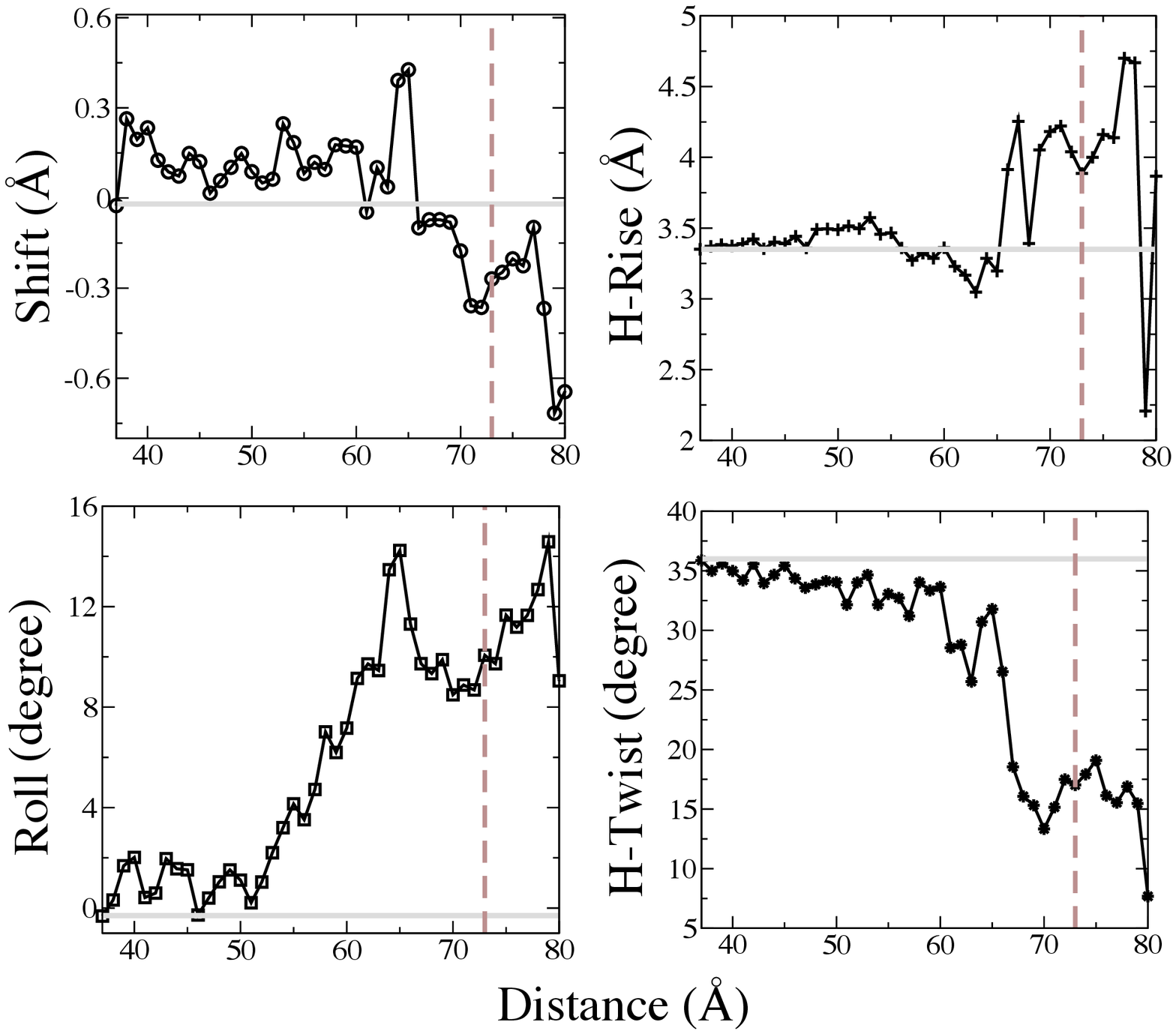}\\
\end{center}
\caption{Average inter base pair parameters of S-DNA are obtained from OS$3$ equilibrium pulling data. Known average inter base pair parameter for B DNA are as follows: Shift=-0.02 {\AA}; Roll=-0.3 degrees; H-Rise=3.35 {\AA}; H-Twist=36.0 degrees [Ref.: R. Lavery {\it et al.}, Nucleic Acids Res. \textbf {37}, 5917 (2009).] and are represented by a solid horizontal line in each plot for reference. Right side of the vertical dotted line (corresponds to $ \sim 140$ pN, obtained from equilibrium stretching curve) in each plot refers to the helical parameters of the emerging S-DNA.}
\label{figs2-hp-inter}
\end{figure}

\begin{figure}
\begin{center}
\includegraphics[width=0.95\columnwidth]{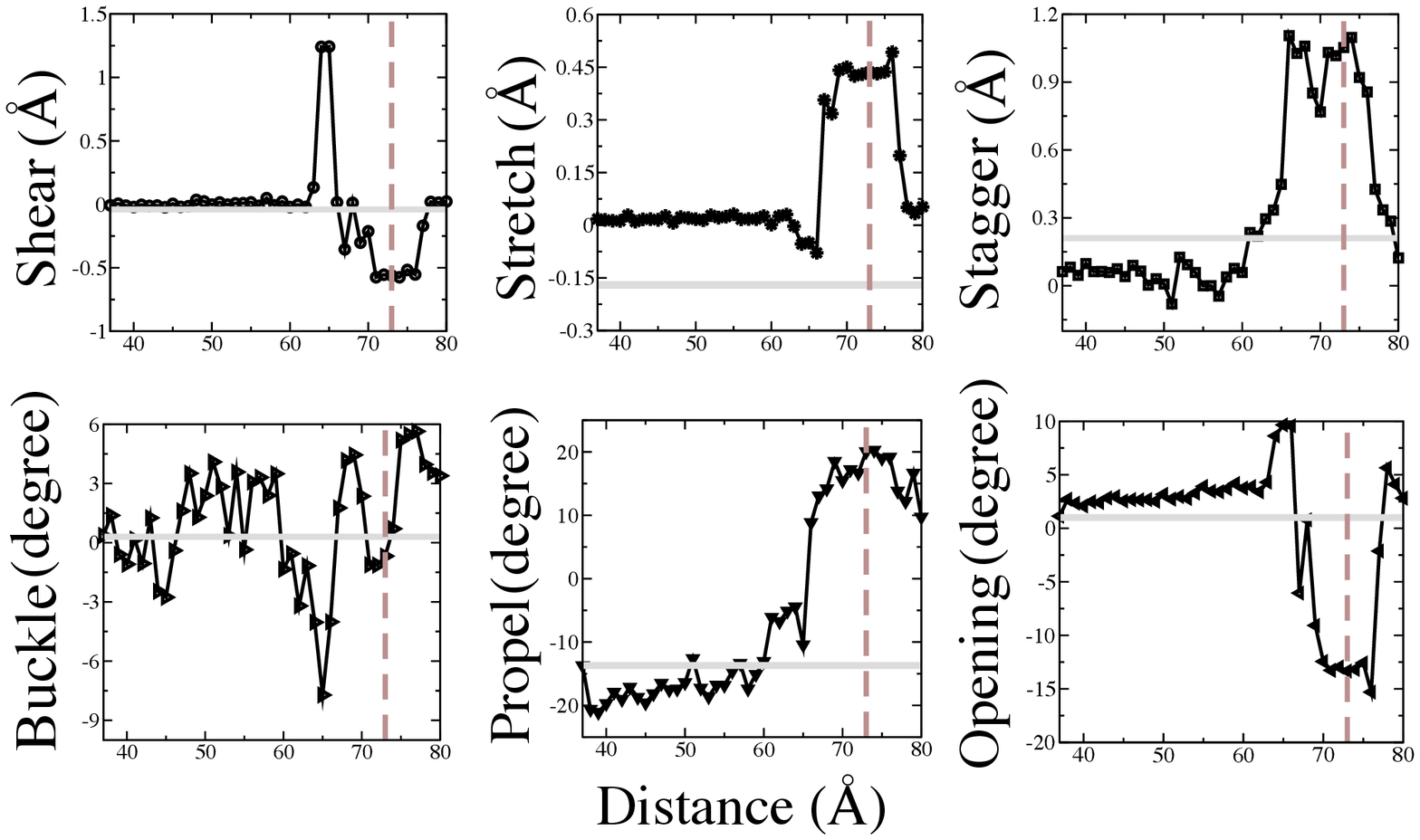}\\
\end{center}
\caption{Average intra base pair parameters of S DNA obtained using Curves+ from OS$3$ equilibrium pulling data. Known average Intra base pair parameters for B DNA are as follows: Shear=-0.04 {\AA}; Stretch=-0.17 {\AA}; Stagger=0.21 {\AA}; Buckle=0.3 degress; Propel=-13.7 degrees; Opening=1.0 degrees [Ref.: R. Lavery {\it et al.}, Nucleic Acids Res. \textbf {37}, 5917 (2009).] and are represented by a solid horizontal line in each plot for reference. Right side of the vertical dotted line (corresponds to $ \sim 140$ pN, obtained from equilibrium stretching curve) in each plot refers to the helical parameters of the emerging S-DNA.}
\label{figs3-hp-intra}
\end{figure}
\begin{figure}
\begin{center}
\includegraphics[width=0.98\columnwidth]{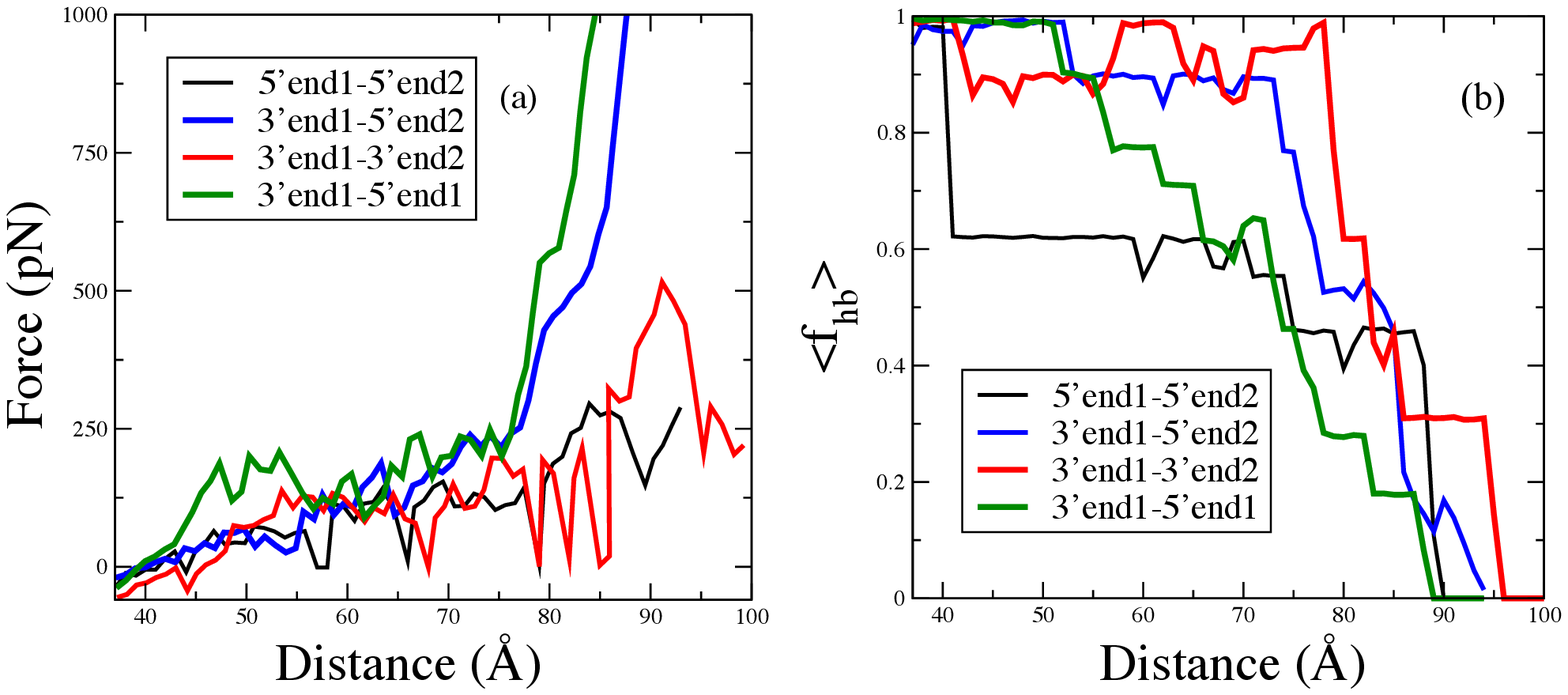}\\
\end{center}
\caption{(a) Force-distance calculated using $F=-k \Delta x$ from equilibrium simulation data at $300$K. (b) Fraction of H-bonds is calculated from equilibrium simulations data at $300$K.} 
\label{figs4-fx-fhb}
\end{figure}
\begin{figure}
\begin{center}
\includegraphics[width=0.95\columnwidth]{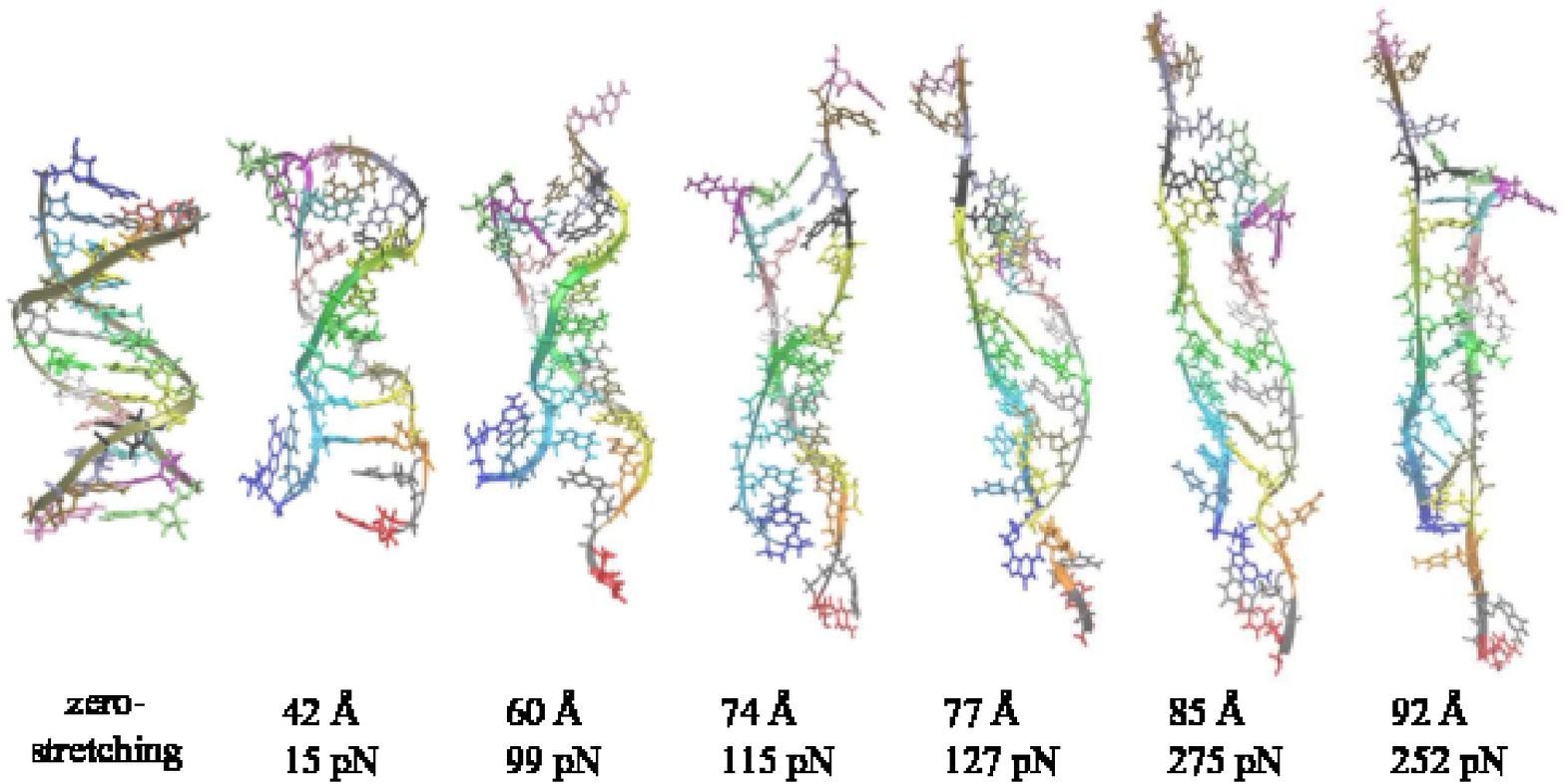}\\
\end{center}
\caption{A few snapshots from our equilibrium simulations for OS$5$ pulling at the $1$-st ns of $12$-bp DNA. Such equilibrium pulling data shows no S-DNA, but small bubbles can be noticed.}
\label{figs5-55-us}
\end{figure}

\begin{figure}
\begin{center}
\includegraphics[width=0.95\columnwidth]{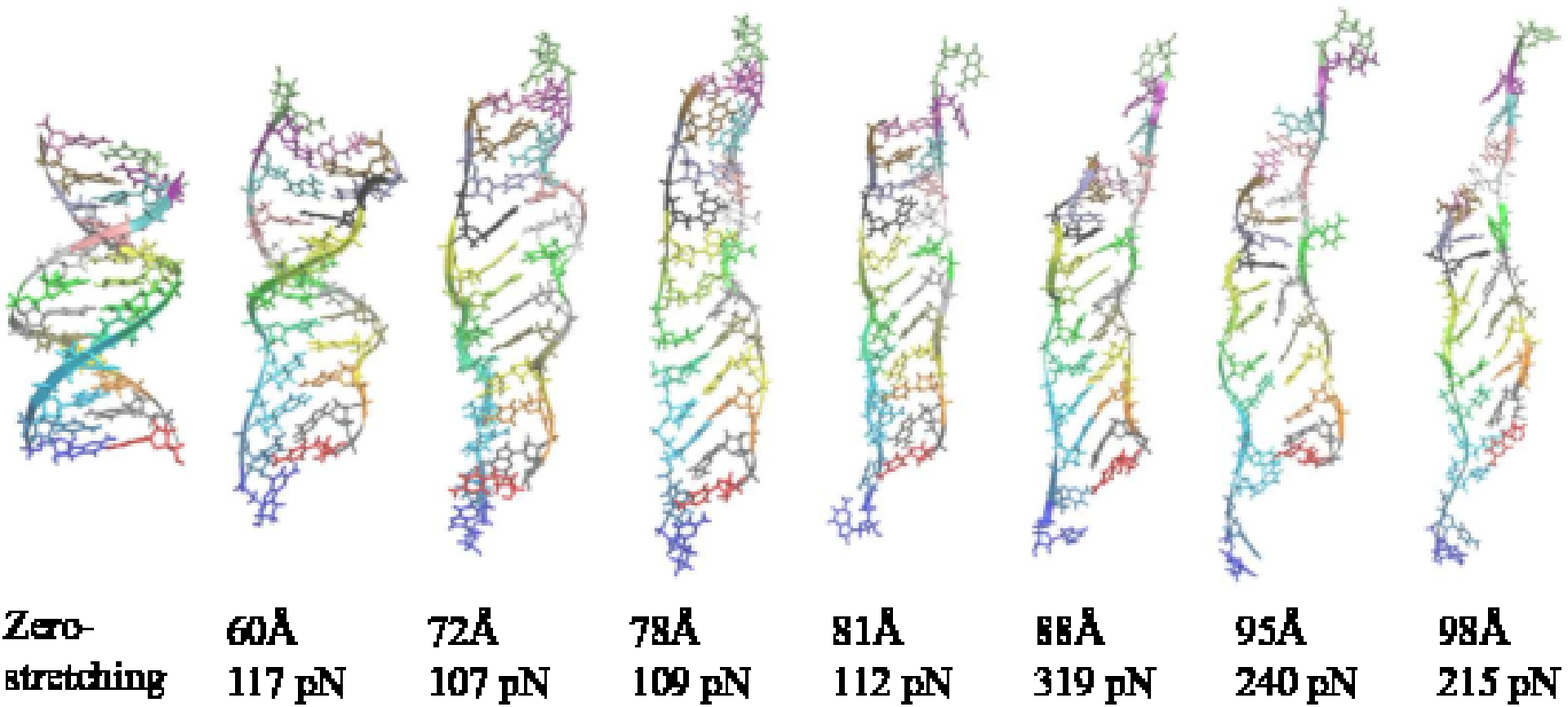}\\
\end{center}
\caption{A few snapshots from our equilibrium simulations for OS$3$ pulling at the $1$-st ns of $12$-bp DNA. Such OS$3$ equilibrium pulling data clearly shows S-DNA.}
\label{figs6-33-us}
\end{figure}
\begin{figure}
\begin{center}
\includegraphics[width=0.95\columnwidth]{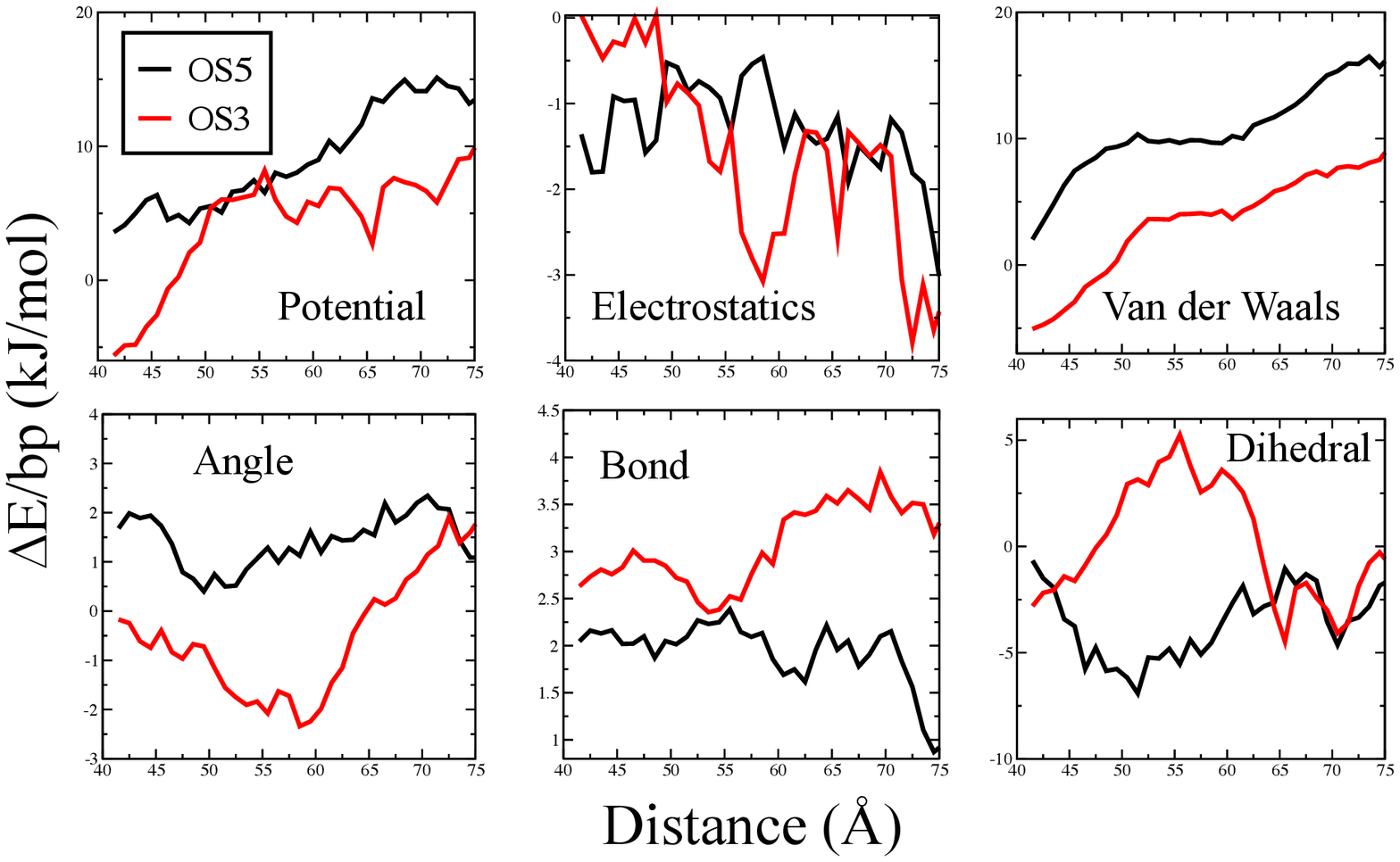}\\
\end{center}
\caption{Different components of the change in the potential of the DNA during OS5 and OS3 pulling.}
\label{figs7-33-55-us}
\end{figure}

\begin{figure}
\begin{center}
\includegraphics[width=0.95\columnwidth]{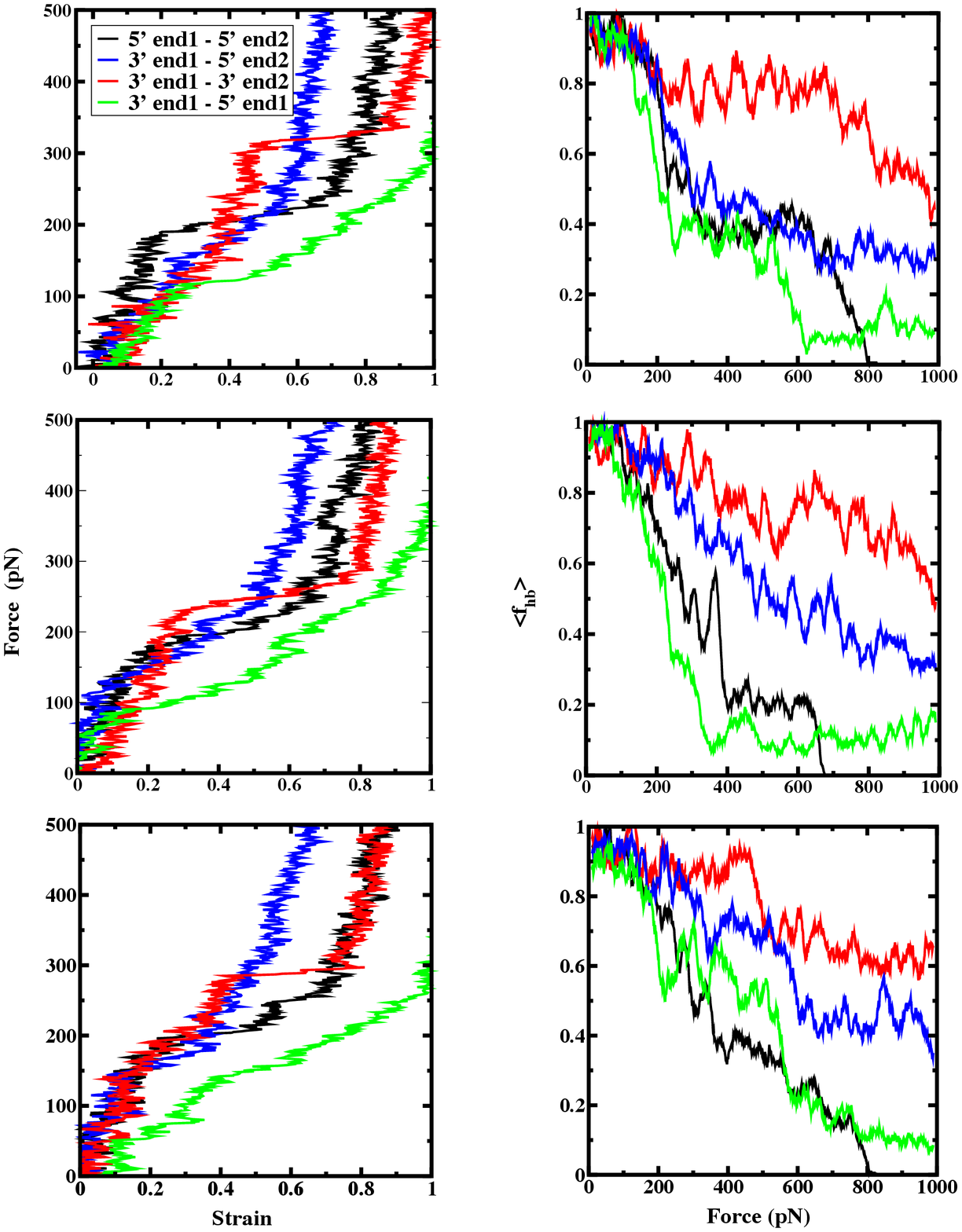}\\
\end{center}
\caption{Force extension behavior and fraction of hydrogen bond $(<f_{hb}>)$ as a function of applied force for various pulling protocols for three independent runs at a pulling rate of $10^{-4}$ pN/fs for $12$bp DNA.}
\label{figs8-all-nep}
\end{figure}
\section{Convergence of PMF calculations}
\begin{figure}
\begin{center}
\includegraphics[width=0.95\columnwidth]{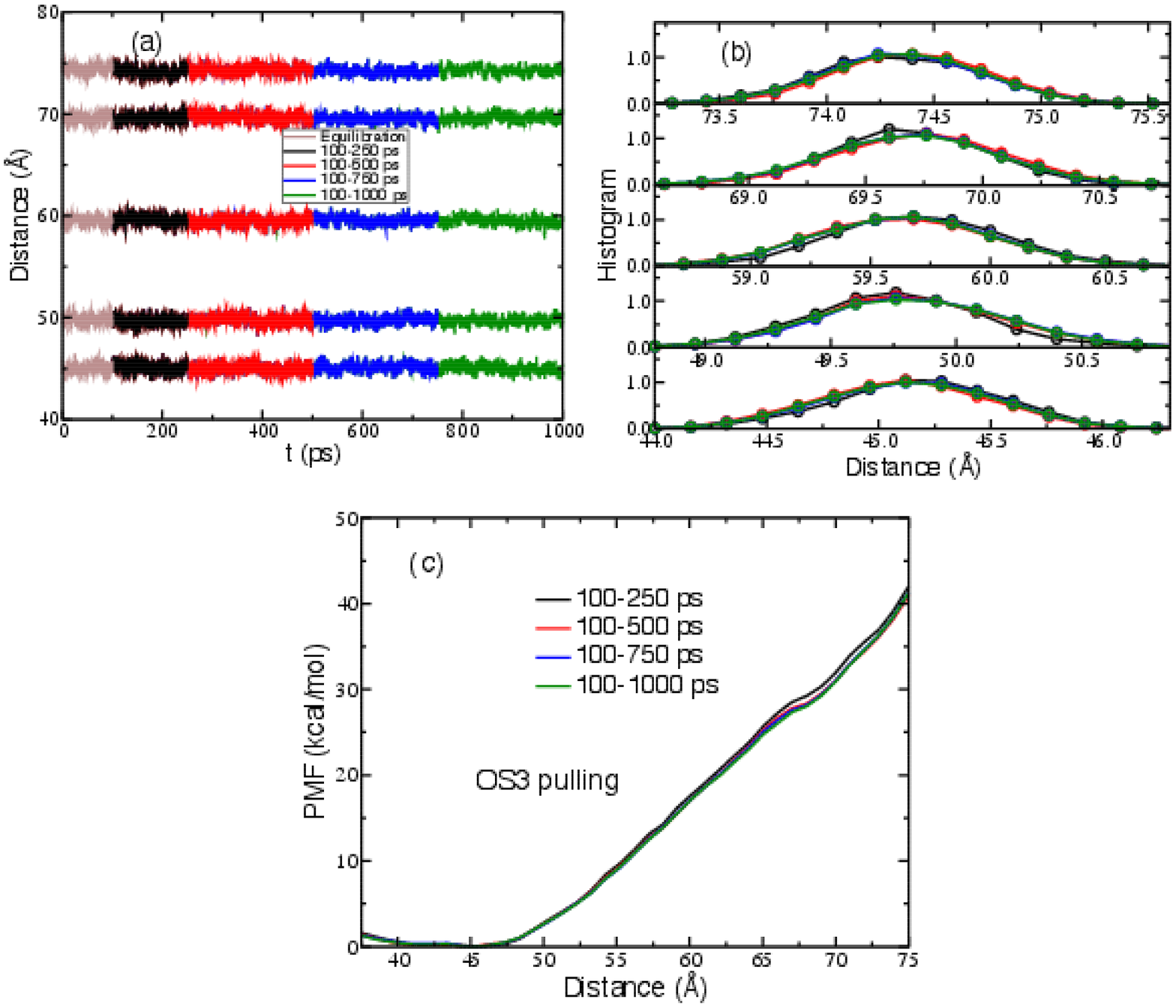}\\
\end{center}
\caption{Convergence test of the PMF calculations for OS3 pulling: (a) Time evolution and (b) corresponding histograms of instantaneous separation between the force centers for selected values of $r_0$ in the restraining potential: $V(r)=k(r-r_0 )^2$. (c) potential of the mean force (PMF) calculated using different lengths of the equilibrium trajectory.}
\label{figs9-pmf-os3-conv}
\end{figure}

We have performed PMF simulations of more than $64$ windows separated by $1$ {\AA} for a time duration of $1$ nanosecond. Due to the backbone torsion potential and base-pair stacking interactions in the helical DNA, achieving the convergence is a quite involved, especially for arbitrary values of restraining spring constant and separation distance between successive PMF windows (Mogurampelly and Maiti JCP 138(3), 034901 (2013)). Therefore, we have initially performed an optimization of spring constant of the restraining potential and the separation distance with an objective of sufficient sampling and overlapping of successive windows. For the optimized values of spring constant ($4$ kcal/mol-{\AA}$^2$) and separation distances ($1$ {\AA}), we subsequently considered the convergence of PMF calculations by sampling the instantaneous distance between the force centers for different lengths of the trajectory. As shown in Figure \ref{figs9-pmf-os3-conv}(a), the instantaneous separation between the force centers (only selected equilibrium distances are presented in the figure) is largely unaffected by the length of simulation. Further, the histograms calculated by sampling the instantaneous distance from different lengths of the trajectory reveal that $500$ ps long trajectory may be sufficient for converging the PMF calculations (see Figure \ref{figs9-pmf-os3-conv}(b)). Consistent with the above picture, the changes in PMF (Figure \ref{figs9-pmf-os3-conv}(c)) are seen to be marginal when the histograms are sampled for $500$ ps or more. Overall, the results presented in Figures \ref{figs9-pmf-os3-conv}(b) and (c) suggest that the PMF calculations are barely affected by the length of the sampling trajectory beyond $500$ ps and that the results of PMF calculation presented in the paper using $1$ ns long trajectory are sufficiently converged.
\begin{figure}
\begin{center}
\includegraphics[width=0.95\columnwidth]{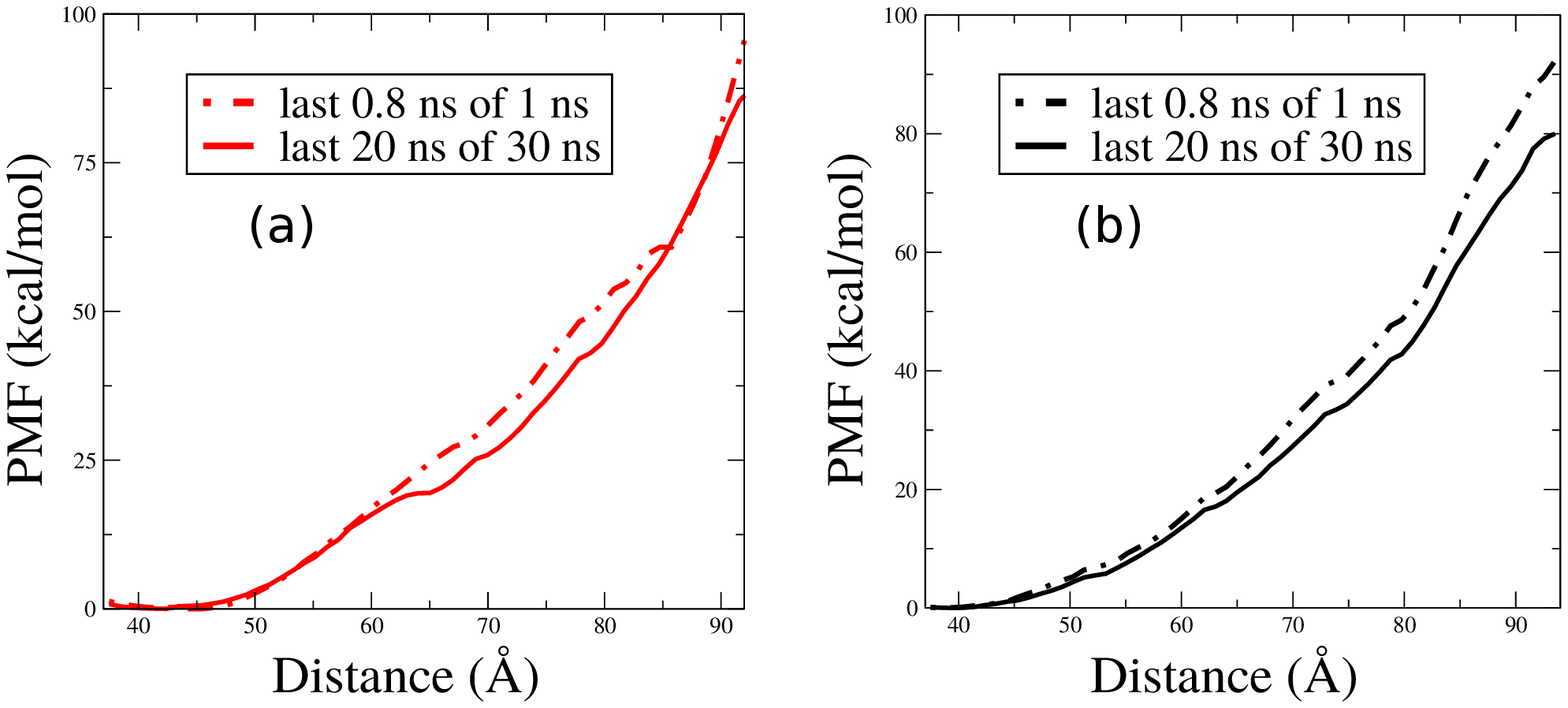}\\
\end{center}
\caption{PMF calculated for (a) OS$3$ and (b) OS$5$ cases and compared between $1$ ns and $30$ ns sampling.}
\label{figs11-33-nep}
\end{figure}

\begin{figure}
\begin{center}
\includegraphics[width=0.95\columnwidth]{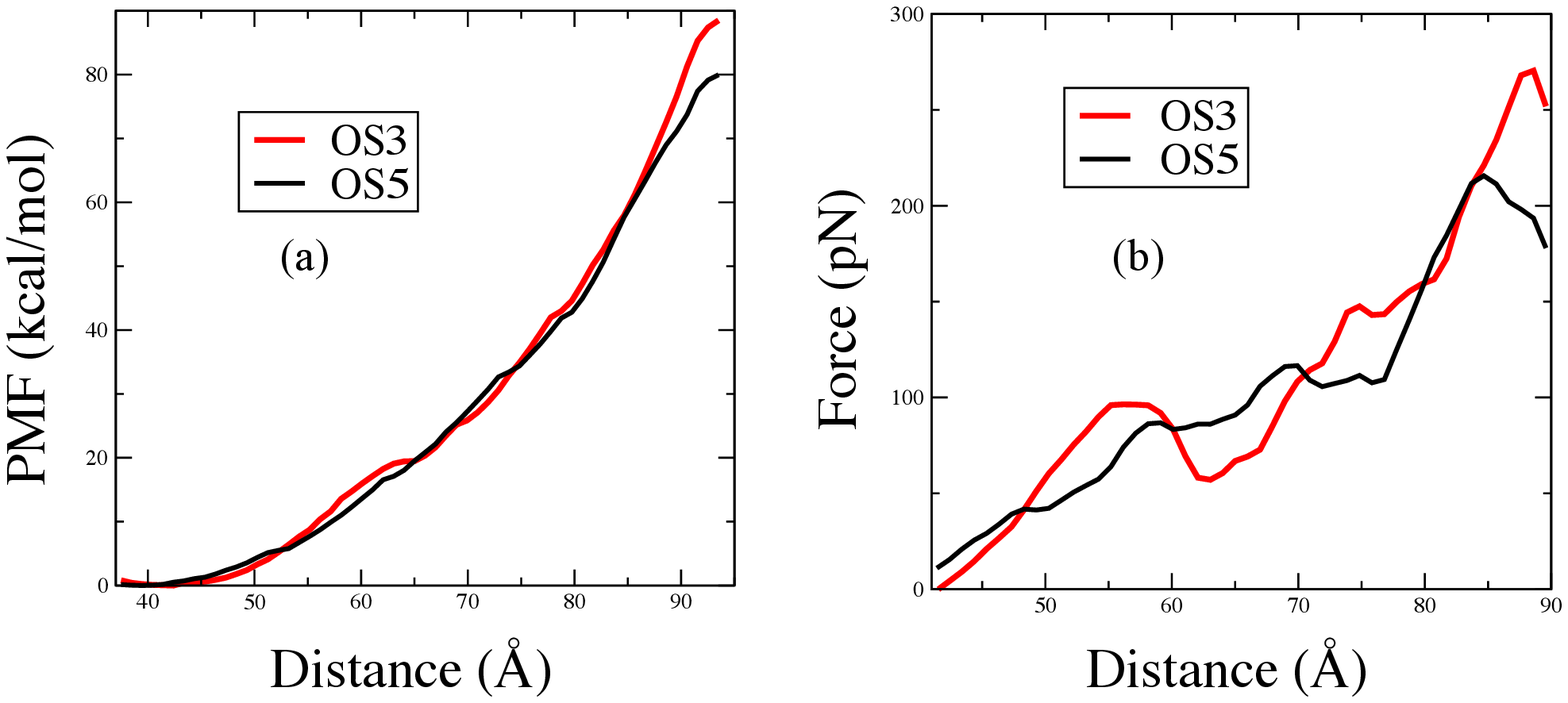}\\
\end{center}
\caption{(a) PMF calculated for OS$3$ and OS$5$ cases while the sampling is done for $30$ ns in each distance window. (b) Force-distance calculated using $F=-k \Delta x$ from these equilibrium simulation data at $300$K. }
\label{figs11-33-nep}
\end{figure}


\begin{figure}
\begin{center}
\includegraphics[width=0.95\columnwidth]{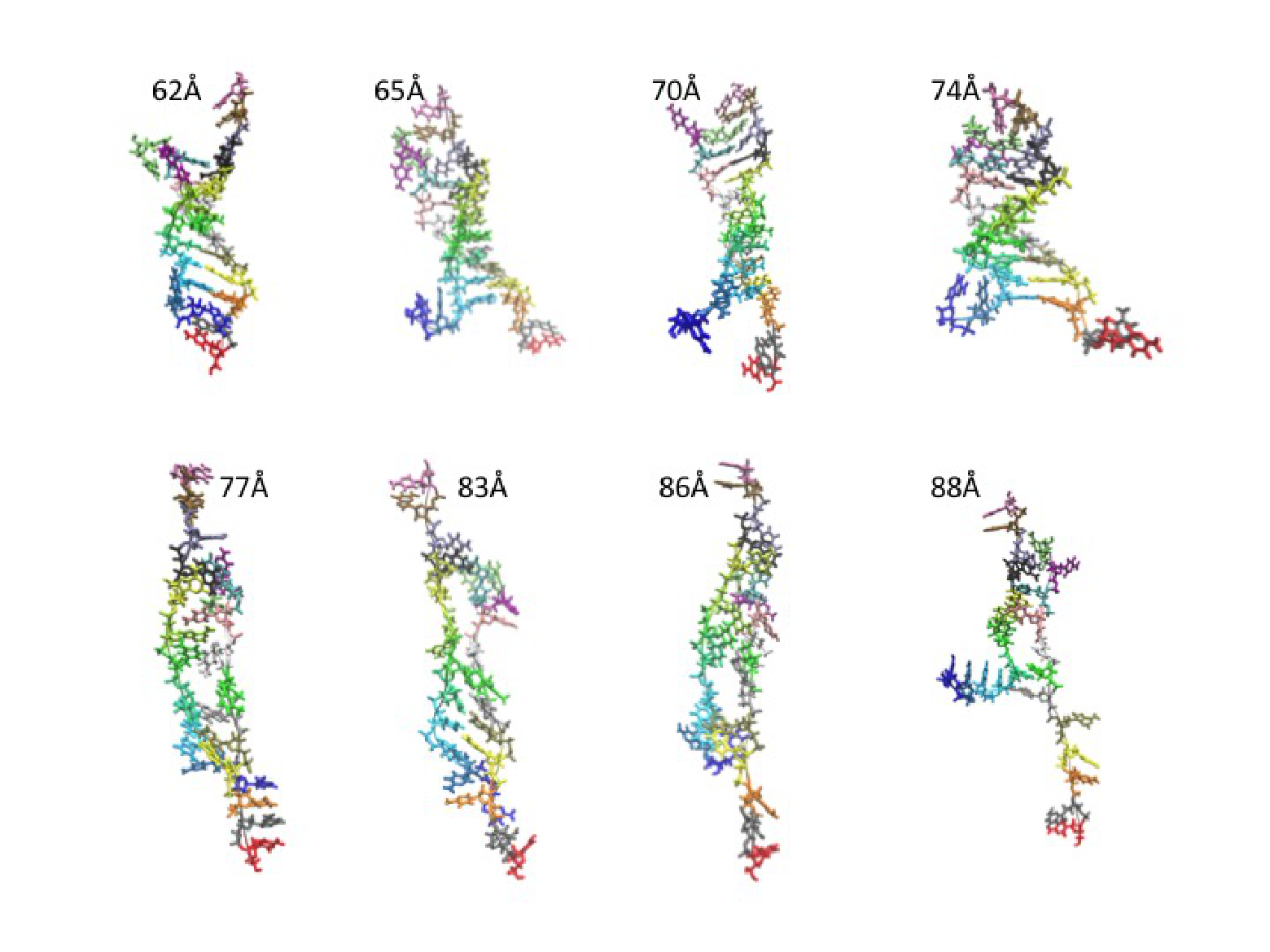}\\
\end{center}
\caption{A few snapshots from our equilibrium simulations at the $30$-th ns for OS$5$ pulling of $12$-bp DNA. Such equilibrium pulling data shows no S-DNA, but small bubbles can be noticed.}
\label{figs12-os5}
\end{figure}
\begin{figure}
\begin{center}
\includegraphics[width=0.95\columnwidth]{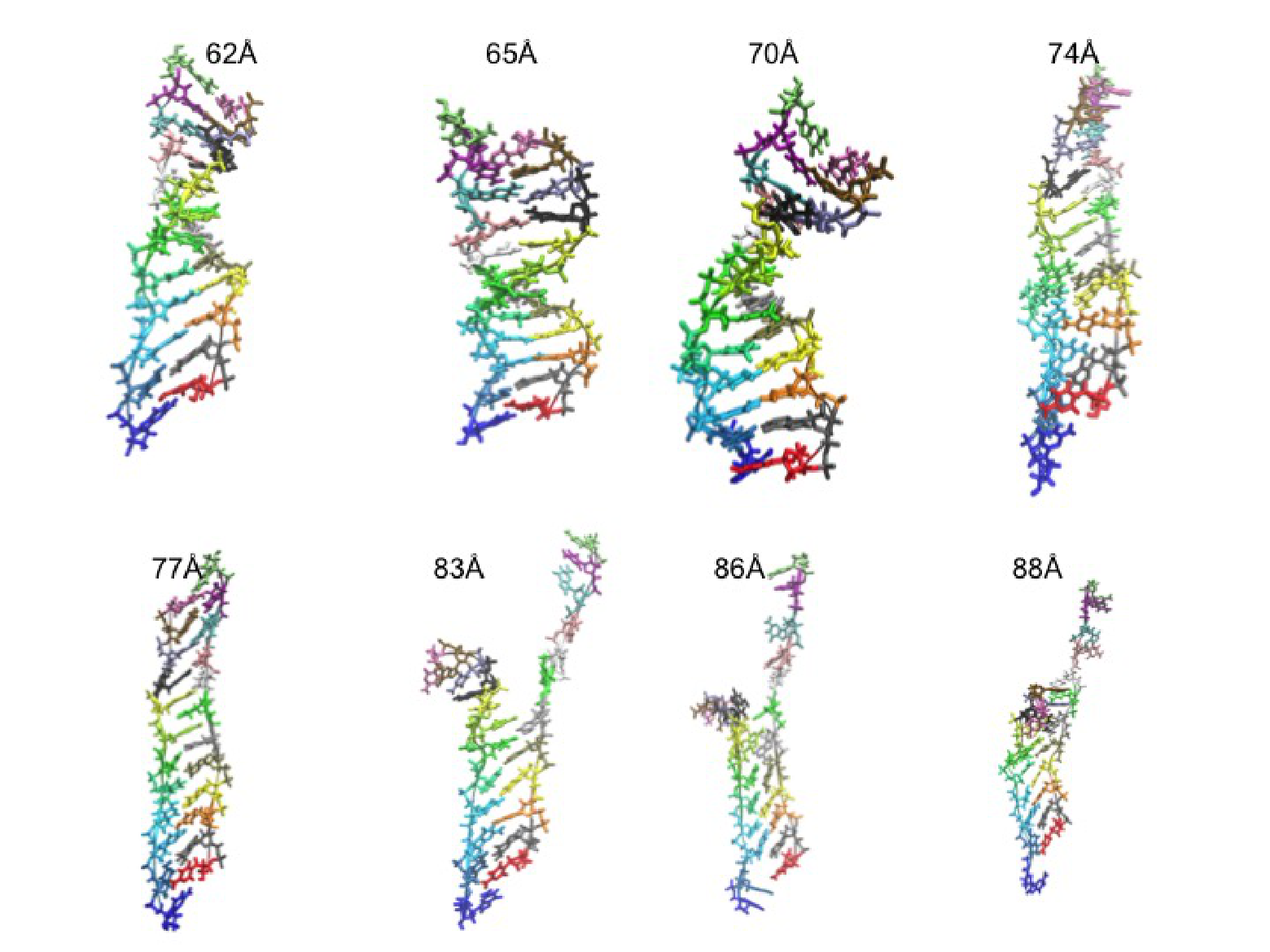}\\
\end{center}
\caption{A few snapshots from our equilibrium simulations at the $30$-th ns  for OS$3$ pulling of $12$-bp DNA. Such OS$3$ equilibrium pulling data clearly shows emergence of S-DNA.}
\label{figs13-os3}
\end{figure}

\begin{figure}
\begin{center}
\includegraphics[width=0.95\columnwidth]{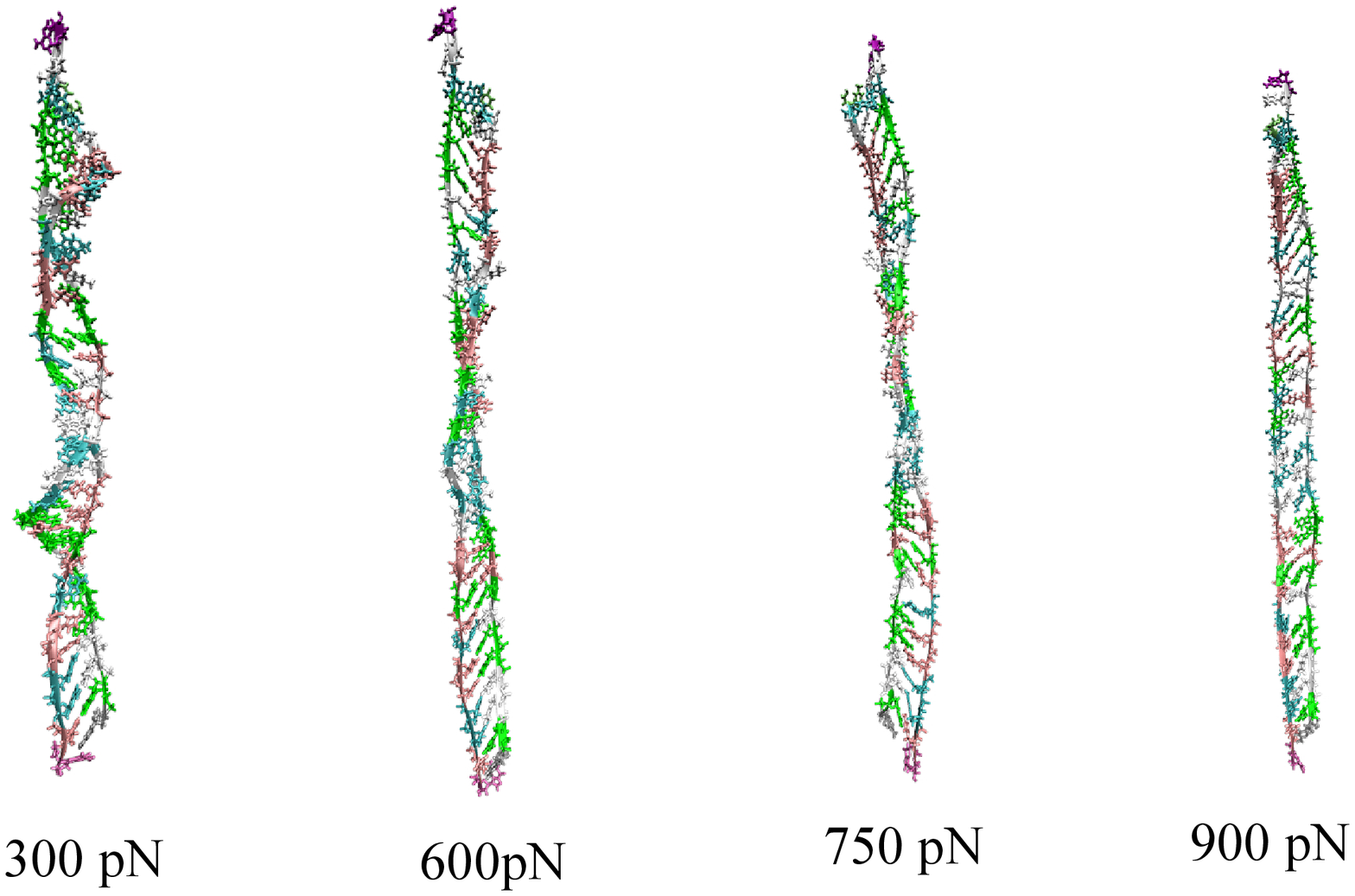}\\
\end{center}
\caption{A few snapshots from our non-equilibrium simulations for OS$3$ pulling for a $30$-bp DNA at different pulling forces. Such non-equilibrium pulling data show S-DNA at $600$ pN, $750$ pN, $900$ pN pulling forces.}
\label{figs14-33-nep}
\end{figure}